\documentclass[final]{article}
\usepackage{times}
\usepackage{a4}
\usepackage{amsmath,amssymb,amsfonts}
\usepackage{fancyvrb}
\usepackage{mathrsfs}
\usepackage[mathcal]{euscript}
\usepackage{amsthm}
\usepackage{stmaryrd}
\usepackage[english]{babel}
\usepackage{graphicx}
\usepackage[all]{xy}

\newcommand{\f}{\ar@{-}} 
\newcommand{\ff}{\ar@{->}} 
\newcommand{\fff}{\ar@{.>}} 
\newcommand{\ffc}{\ar@/^/} 
\newcommand{\fffc}{\ar@/_/} 
\newcommand{\fffp}{\ar@/./} 

\theoremstyle{plain}
\newtheorem{theorem}{Theorem}[section] 

\newtheorem{corollary}[theorem]{Corollary}
\newtheorem{lemma}[theorem]{Lemma}
\newtheorem{remark}[theorem]{Remark}
\theoremstyle{definition}
\newtheorem{definition}[theorem]{Definition}
\newtheorem{example}[theorem]{Example}
\newtheorem{hypothesis}[theorem]{Hypothesis}

\newcommand{\rmucalculusstar}{\mbox{\raisebox{-0.25ex}{$\stackrel{\mbox{\tiny$\curvearrowleft$}}{\mu\mskip-3mu\smash{{}^{\scriptscriptstyle\star}}\mskip1mu}$}}}

\newcommand{\CTLstar}{{\mathrm{CTL}^*}}

\newcommand{\ltl}{\mathrm{LTL}}

\newcommand{\ctls}{\mathrm{CTL}^*}
\newcommand{\ctl}{\mathrm{CTL}}

\newcommand{\actl}{\mathrm{ACTL}}

\newcommand{\ltldet}{\mathrm{LTL}_\mathrm{det}}

\newcommand{\U}{\mathrm{U}}

\newcommand{\W}{\mathrm{W}}

\newcommand{\integer}{{\mathbb{Z}}}

\newcommand{\comment} [1]{}

\newcommand{\bneg}{\boldsymbol{\neg}}
\newcommand{\id}{\mathrm{id}}
\newcommand{\bZ}{\mathbb{Z}}
\newcommand{\boplus}{{\boldsymbol{\oplus}}}
\newcommand{\bominus}{{\boldsymbol{\ominus}}}

\DeclareMathOperator{\traces}{{\mathit{Traces}}}
\DeclareMathOperator{\States}{{\mathit{States}}}
\DeclareMathOperator{\states}{{\mathit{States}}}
\DeclareMathOperator{\MC}{MC}
\DeclareMathOperator{\Cl}{Cl}
\DeclareMathOperator{\Shell}{Shell}
\DeclareMathOperator{\shell}{Shell}

\DeclareMathOperator{\core}{Core}
\DeclareMathOperator{\pre}{pre}
\DeclareMathOperator{\post}{post}
\DeclareMathOperator{\uco}{uco}

\DeclareMathOperator{\lfp}{lfp}
\DeclareMathOperator{\gfp}{gfp}

\DeclareMathOperator{\img}{img}

\newcommand{\pret}{\ensuremath{\widetilde{\pre}}}
\newcommand{\postt}{\ensuremath{\widetilde{\post}}}
\newcommand{\pr}{\ensuremath{P_{{\scriptscriptstyle \!\!\rightarrow}}\!}}

\newcommand{\zm}{{\ok{\mathbb{Z}_{\leq 0}}}}

\def \tuple#1{\langle #1 \rangle}
\newcommand{\ddef}{\hfill$\Box$}
\newcommand{\ud}{\mbox{\raisebox{0ex}[1ex][1ex]{$\:\stackrel{{\scriptscriptstyle 
\mathrm{def}}}{=}\:$}}}
\newcommand{\nat}{\mathbb{N}}
\newcommand*{\gdca}{(\alpha ,C,A,\gamma)}

\newcommand{\rak}{\ok{\stackrel{\scriptscriptstyle k}{\rightarrow}}}
\newcommand{\sv}{{\ok{{\wp (\mathbb{S})}^{_{^{\!\stackrel{\shortleftarrow}{\omega}}}}}}}
\newcommand{\svo}{{\ok{{\wp (\mathbb{S})}^{\omega}}}}

\def\rakn#1{\ok{\stackrel{\scriptscriptstyle #1}{\rightarrow}}}
\def\grass#1{[\![#1]\!]}
\def\grasse#1{[\![#1]\!]}
\def\grassg#1{\{\!\hspace*{-.6pt} | #1| \hspace*{-.6pt}\! \}}

\def\cM{\mathcal{M}}

\def\cA{\mathcal{A}}

\newcommand{\ra}{\rightarrow}
\newcommand{\sra}{\shortrightarrow}
\newcommand{\da}{\downarrow}
\newcommand{\Ra}{\Rightarrow}
\def\rat{\!\shortrightarrow\!}

\def\fab{\ok{\boldsymbol{\forall}}}

\def\cal{{\curvearrowleft}}
\def\cl{\ok{\mbox{}^\curvearrowleft}}
\def\mus{\ok{\rmucalculusstar}}

\def\clm{\ok{\mathfrak{L}_{\scriptstyle{\stackrel{{\curvearrowleft_{\!\!\!\star}}}{\mu}}}}}

\def\mtr{{\mathscr{M}_{\shortrightarrow}}}
\def\ma{{\mathscr{M}_{\shortrightarrow}}}

\def\gabba{{\square \!\!\!\! \mbox{}^{\mbox{}_{\mbox{}_{\scriptscriptstyle \pm}}}}}

\def\rfm{{\ok{\rho^\forall_M}}}

\def\bull2{\mbox{\begin{picture}(1,1)(1,1)\circle*{1} \end{picture}}}

\def\Lra{\Leftrightarrow}

\def\ok#1{\mbox{\raisebox{0ex}[1ex][1ex]{$#1$}}}
 
\begin{document}

\title{{\bf \Large {Incompleteness of States w.r.t.\ Traces\\ in Model Checking}}}

\author{
\normalsize {\sc Roberto Giacobazzi} \\
\small Dipartimento di Informatica\\ 
\small Universit\`a di Verona, Italy\\
\small \texttt{roberto.giacobazzi$@$univr.it}
\and 
\normalsize {\sc Francesco Ranzato}
\\
\small Dipartimento di Matematica Pura ed Applicata\\ 
\small Universit\`a di Padova, Italy\\
\small \texttt{francesco.ranzato$@$unipd.it}
}

\date{}
\pagestyle{plain}

\maketitle

\begin{abstract}
Cousot and Cousot introduced and studied a general
past/future-time specification language, called $\mus$-calculus,
featuring a natural time-symmetric \emph{trace-based} semantics. The
standard  \emph{state-based} semantics of the $\mus$-calculus is an
\emph{abstract interpretation} of its trace-based semantics, which 
turns out to be \emph{incomplete} (i.e., trace-incom\-plete), even for finite
systems. As a consequence, standard state-based model checking of the
$\mus$-calculus is
incomplete w.r.t.\ trace-based model checking. 
This paper shows that any refinement or abstraction of the domain of
sets of 
states induces a  corresponding semantics which is still
trace-incomplete for any propositional fragment of 
the $\mus$-calculus. This derives from 
a number of results, one for each incomplete logical/temporal
connective of the $\mus$-calculus, that characterize the structure of
models, i.e.\ transition systems, whose corresponding
state-based semantics of the $\mus$-calculus is trace-complete. 
\end{abstract}

\section{Introduction}\label{intro}

Temporal specification languages used in automatic verification by
model checking can be classified in two broad classes: linear and
branching time languages. Linear-time languages allow to express
properties of computation paths of the model, called traces, while
specifications of branching time languages describe properties that
depend on the branching structure of the model.  $\ltl$ and $\ctl$ are
the most commonly used languages for, respectively, linear and
branching time model checking.  The relationship between linear and
branching time languages has been the subject of thorough
investigation since the 1980s (see \cite{var01} for a survey), in
particular it is well known that $\ltl$ and $\ctl$ have incomparable
expressive powers \cite{cd88,eh86,lam80}.

Given a linear specification $\phi$, the standard universal model
checking problem consists in characterizing the set $\MC_M^\forall(
\phi)$ of states $s$ in a model $M$, i.e.\ a transition system (or a
Kripke structure), such that any trace in $M$ whose present time is
$s$ satisfies $\phi$. Hence, if $\grasse{\phi}=\{ \tuple{i,\sigma} \in
M~|~ \tuple{i,\sigma} \models \phi\}$ denotes the trace semantics of
$\phi$, where in a trace $\tuple{i,\sigma}$, $\sigma$ is a
$\mathbb{Z}$-indexed sequence of states and $i\in \mathbb{Z}$ denotes
present time, then $\MC_M^\forall (\phi) = \{s\in \States~|~ \forall
\tuple{i,\sigma} \in M.\, (\sigma_i=s) \Ra \tuple{i,\sigma} \in
\grasse{\phi}\}$.  Cousot and Cousot showed in their POPL'00 paper
\cite{CC00} that this can be formalized as a step of abstraction
within the standard abstract interpretation framework
\cite{CC77,CC79}. In fact, Cousot and Cousot \cite{CC00} consider the
universal path quantifier $\alpha_M^\forall: \wp(\traces)\ra
\wp(\States)$ which maps any set $T$ of traces to the set of states
$s\in \States$ such that any trace in $M$ with present state $s$
belongs to $T$ and show that $\alpha_M^\forall$ is an approximation
map in the abstract interpretation sense. Hence, $\alpha_M^\forall$ is
called the \emph{universal model checking abstraction} because
$\MC_M^\forall (\phi) = \alpha_M^\forall (\grasse{\phi})$.  Dually,
one can define an \emph{existential} model checking abstraction
$\alpha_M^\exists: \wp(\traces)\ra \wp(\States)$ that formalizes
standard existential model checking: $\alpha_M^\exists(T)$ provides
the set of states $s\in \States$ such that there exists a trace in $M$
with present state $s$ which belongs to $T$. According to the standard
abstract interpretation methodology, this universal abstraction gives
rise to an abstract state semantics of a linear language and thus
transforms the trace-based universal model checking problem to a
state-based universal model checking problem. Basically, the universal
state-based semantics $\grasse{\phi}_{\mathrm{state}}^\forall$ of a
linear formula $\phi$ is obtained by abstracting each linear temporal
operator appearing in $\phi$, like next-time or sometime operators, to
its best correct approximation on $\wp(\mathit{States})$ through the
abstraction map $\alpha_M^\forall$.  This abstract semantics
$\grasse{\phi}_{\mathrm{state}}^\forall$ of $\phi$ coincides with the
state semantics of the branching time formula $\phi_\forall$ obtained
from $\phi$ by preceding each linear temporal operator occurring in
$\phi$ by the universal path quantifier.  In Cousot and Cousot's work
\cite{CC00} formulae range over a past- and future-time temporal
language which generalize Kozen's $\mu$-calculus.  Hence, this allows
to transform the trace-based model checking problem $M,s
\models_{\mathrm{trace}} \phi$, i.e.\ $s\in \alpha_M^\forall (\grasse{
  \phi})$, to a state-based model checking problem $M,s
\models_{\mathrm{state}} \phi$, i.e.\ $s\in
\grasse{\phi}_{\mathrm{state}}^\forall$.
 
It should be clear that the state-based model checking is a
\emph{sound} approximation of the trace-based one, namely:
$$M,s \models_{\mathrm{state}} \phi \; \Ra \;  M,s
\models_{\mathrm{trace}} \phi.$$ It should be noted that in abstract
interpretation soundness is guaranteed by construction, namely
$\grasse{\phi}_{\mathrm{state}}^\forall\subseteq \alpha_M^\forall
(\grasse{ \phi})$ always holds.  However, it turns out that this
abstraction is \emph{incomplete}, that is, the reverse direction does
not hold, even for finite-state systems. We will provide later an
example for this phenomenon.  Let us remark that when
$\grasse{\phi}_{\mathrm{state}}^\forall = \alpha_M^\forall (\grasse{
  \phi})$ holds for some linear formula $\phi$, Kupferman and
Vardi~\cite{kv98,var98} say that the formula $\phi$ is
\emph{branchable}.  Branchable formulae have been used by Kupferman
and Vardi for studying how model checking of a $\ltl$ formula $\phi$
can be reduced to an equivalent model checking of the corresponding
$\ctl$ formula $\phi_\forall$.

The above incompleteness means that universal model checking of linear
formulae cannot be reduced with no loss of precision to universal
model checking on states through the universal abstraction. This also
means that classical state-based model checking algorithms (e.g.\ for
$\ctl$) do not provide exact information w.r.t.\ a trace-based
interpretation.  This opens the question whether it is possible to
find some different approximation $\cA$ of the trace-based model
checking problem which (1)~is still related to states, namely $\cA$
refines or abstracts from sets of states, and (2)~induces an
approximated model checking which is instead equivalent to the
trace-based one: for any $s\in \states$ and any linear formula $\phi$,
$$M,s\models_{\cA}\phi \; \Lra \; M,s\models_{\mathrm{trace}}
\phi.\eqno(*)$$ It is important to remark that we do not consider
generic approximations of traces, but only approximations that can be
obtained by refinements or simplifications of sets of states, namely
of the domain $\wp(\states)$. Let us notice that the trivial
abstraction $\mathrm{Trivial}\ud \{\bot\}$, i.e.\ the abstraction
carrying no information at all by confusing all the traces, i.e.\
$\alpha_{\mathrm{Trivial}}(T)=\bot$ for any set $T$ of traces,
satisfies the above equivalence because we always have that
$\grasse{\phi}_{\mathrm{Trivial}} = \bot = \alpha_{\mathrm{Trivial}}
(\grasse{ \phi})$.  More precisely, the paper answers the following
question: is it possible to minimally refine or abstract the
state-based semantics of a general temporal languages so that this
refinement/abstraction induces a corresponding approximated model
checking which is trace-complete, i.e.\ equivalent to the trace-based
model checking? In our approach, refinements and abstractions of a
semantics are intended to be specified by \emph{standard abstract
  interpretation} \cite{CC77,CC79}.  This paper provides the following
results:
\begin{itemize}
\item[{\rm (i)}] the only
refinement of the state-based semantics inducing a trace-complete
model checking is the trace-based semantics itself;
\item[{\rm (ii)}] on the opposite direction, the only
abstraction of the state-based semantics inducing a trace-complete
model checking is the trivial semantics carrying no information at
all;
\item[{\rm (iii)}] for each basic temporal/logical operator of a past- and
future-time extension of Kozen's
  $\mu$-calculus we characterize the least trace-complete abstractions
  which, respectively, include and are included in the state-based
semantics.
\end{itemize}
Points {\rm (i)} and {\rm (ii)} prove that states are, so to say,
``intrinsically trace-incomplete'', since there is no way to obtain a
trace-complete model checking by modifying, through refinements or 
abstractions, the state-based semantics.

\paragraph*{\textbf{The Scenario.}}
As mentioned above, our results are formulated and shown within the
Cousot and Cousot's~\cite{CC00} abstract interpretation-based approach
to model checking called \textit{temporal abstract
  interpretation}. Cousot and Cousot~\cite{CC00} introduced an
enhanced past- and future-time temporal calculus, called
$\mus$-calculus, which is inspired by Kozen's $\mu$-calculus.  The
trace-based semantics of the $\mus$-calculus is time-symmetric: this
means that execution traces have potentially infinite length both in
the future and in the past.  This time symmetry is not the only
feature of the $\mus$-calculus. The $\mus$-calculus also provides a
tight combination of linear and branching time, allowing to derive
classical specification languages like $\ltl$, $\ctl$, $\ctls$ and
Kozen's $\mu$-calculus itself, as suitable fragments.

One main achievement in \cite{CC00} is that state-based model checking
of transition systems (or Kripke structures) can be viewed as an
abstract interpretation of the trace-based semantics. It is worth
mentioning that this abstract interpretation-based approach has been
applied to a number of temporal languages by Schmidt~\cite{sch01} and
also to the case of modal Kripke transition systems by
Schmidt~\cite{sch01} and Huth et al.~\cite{hjs01}.  The semantics
$\ok{\grass{\phi}_\mathrm{trace}}$ of a temporal specification
$\phi\in \mus$ is the set of traces in the model $M$ making $\phi$
true.  States are viewed as a universal abstract interpretation of
traces through the universal concretization
$\gamma_M^\forall:\wp(\states)_{\supseteq} \ra
\wp(\traces)_{\supseteq}$ defined by $$\gamma_M^\forall
(S)=\{\tuple{i,\sigma}\in M~|~\sigma_i\in S\}.$$ This maps
$\gamma_M^\forall$ induces an abstract interpretation together with
its adjoint universal abstraction $\ok{\alpha_M^\forall}:
\wp(\mathit{Traces}) \ra \wp(\mathit{States})$ defined by
$$\alpha_M^\forall(T)= \{s\in \mathit{S}~|~\text{ for any trace
}\tuple{i,\sigma} \in M, \text{ if } \sigma_i  =s \text{ then
} \tuple{i,\sigma}\in T\}.$$
This abstract interpretation systematically induces
a state-based
semantics $\ok{\grass{\cdot}_\mathrm{state}^\forall}: \mus \ra
\wp(\mathit{States})$: for example, for an atomic proposition $p$, 
\begin{align*}
\ok{\grass{p}_\mathrm{state}^\forall} & \ud \ok{\alpha_M^\forall
(\grass{p}_\mathrm{trace})}\\
\ok{\grass{\mathrm{AX}p}_\mathrm{state}^\forall} & \ud \ok{\alpha_M^\forall \circ
\mathbf{X} \circ \gamma_M^\forall}
\ok{(\grass{p}_\mathrm{state}^\forall)} = \ok{\pret_\sra}
\ok{(\grass{p}_\mathrm{state}^\forall)} 
\end{align*}
where $\mathbf{X}$ is the next-time transformer on traces and
$\ok{\pret_\sra}$ is the standard ``universal pre'' transformer of
states w.r.t.\ the transition relation $\sra$ of the model $M$.  The
abstract interpretation approach ensures that
$\ok{\grass{\cdot}_\mathrm{state}^\forall}$ is sound by construction
with respect to the trace semantics: for any $\phi\in \mus$:
$$\ok{\grass{\phi}_\mathrm{state}^\forall \subseteq
  \alpha_M^\forall (\grass{\phi}_\mathrm{trace})}.$$ However, as
proved in \cite{CC00}, this inclusion may be strict and this means
that the state-based model checking of the $\mus$-calculus is
trace-incomplete, namely the above equivalence~$(*)$ does not
hold. Let us recall an example of incompleteness from \cite{CC00}.

\begin{example}\label{first}
Consider the following
minimal transition system $M$: 
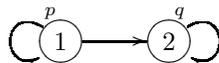
\begin{center}
\mbox{\xymatrix{
     *++[o][F]{1} \ar@(ul,dl)[] _(0.1){p} \ar[r]  & 
*++[o][F]{2} \ar@(ur,dr)[] ^(0.1){q}
\\
}
}
\end{center}
and consider the linear formula $\phi = \mathrm{G} p \vee 
\mathrm{FG} q$. 
We have that
\begin{align*}
\grasse{\mathrm{G}p}_\mathrm{trace} &=
\{\tuple{i,\sigma}\in M~|~\forall j \geq i.\: \tuple{j,\sigma} \in 
\grasse{p}_{\mathrm{trace}}\}
= \{
\tuple{i,\cdots~1~1~1~\cdots}\in M~|~i\in \mathbb{Z}\}\\
\grasse{\mathrm{FG}q}_\mathrm{trace} &=
\{\tuple{i,\sigma}\in M~|~\exists j\geq i.\: \forall k\geq j.\:
\tuple{k,\sigma} \in \grasse{p}_{\mathrm{trace}}\}\\
& = \{
\tuple{i,\cdots~1~1~1~2~2~2\cdots}\in M~|~i\in \mathbb{Z}\} \cup 
\{
\tuple{i,\cdots~2~2~2~\cdots}\in M~|~i\in \mathbb{Z}\}.
\end{align*}
Thus, $\grasse{\phi}_\mathrm{trace} = M$, so that $\alpha_M^\forall
(\grasse{\phi}_\mathrm{trace}) = \{1,2\}$.  On the other hand, we have
that the state semantics $\grasse{\phi}_\mathrm{state}^\forall$ is
given by the state semantics of the $\ctl$ formula $\phi_\forall =
\mathrm{AG}p \vee \mathrm{AFAG}q$. Thus, it turns out that
$\grasse{\phi}_\mathrm{state}^\forall = \{2\}$ because in $M$: (i) it
is possible to jump from state $1$ to state $2$ so that
$\grasse{\mathrm{AG}p}=\varnothing$ and (ii) it is possible to stay
forever in state $1$ so that $\grasse{\mathrm{AFAG}q}=\{2\}$.  As a
consequence,
$$M,1 \models_{\mathrm{trace}} \phi\text{~~~~while~~~~}
M,1\not\models_{\mathrm{state}}\phi$$
namely, the universal state-based
model checking of state $1$ for $\phi$ is trace-incomplete. 
\ddef
\end{example}

The same phenomenon holds even for standard, i.e.\ partition-based
\cite{CGL94,CGP99}, or generic, i.e.\ abstract domain-based
\cite{CC00,GQ01,RT04,RT05}, abstract model checking where the
abstraction map actually is a state-abstraction and can be modeled as
a further abstract interpretation step of
$\ok{\grass{\cdot}_\mathrm{state}}$.  It is therefore important in
order to understand the limits of state-based (concrete or abstract)
model checking with respect to properties of traces, to investigate
whether it is possible to find a semantics $\ok{\grass{\cdot}_{?}}$ as
a refinement or abstraction of $\ok{\grass{\cdot}_\mathrm{state}}$
which is complete for the trace-based semantics
$\ok{\grass{\cdot}_\mathrm{trace}}$.

\paragraph*{\textbf{Complete Core and Shell.}}
Our main goal is that of isolating
the least refinements and abstractions of the state-based
model checking, i.e.\ of $\wp(\states)$ viewed as 
abstract domain of $\wp(\traces)$ through the universal abstraction
$\alpha_M^\forall$, which are
trace-complete. 

Let us recall that an abstract domain
$A=\alpha(\mathit{Concrete})$ together with an 
abstract semantics $\ok{f^\sharp} :A\ra A$
is \emph{complete} for a semantic function
$f:\mathit{Concrete}\ra \mathit{Concrete}$ when
$\alpha(f(c))=\ok{f^\sharp}(\alpha(c))$ holds for any concrete
$c$. Thus, completeness means
that abstract computations by $\ok{f^\sharp}$ are as
precise as possible in the abstract domain $A$. 
Giacobazzi et al.\ \cite{jacm} observed that
completeness actually depends on the abstract
domain $A$ only, because it is enough to consider 
the best correct approximation $\alpha \circ f \circ \gamma$ of $f$ as
abstract semantics. Thus, it turns out that completeness
is an abstract domain property: $A$ is
complete for $f$ iff the equation $\alpha \circ f  =
\alpha \circ f\circ \gamma \circ \alpha$ holds. 
Hence, this opens up the key question of making an abstract
interpretation complete by minimally extending or restricting the
underlying abstract domain.  Following the terminology in \cite{jacm},
we call \emph{complete shell}/\emph{core} of $A$ the most
abstract/concrete domain, when this exists, which refines/abstracts
$A$ and is complete for $f$. Thus, complete shells add to an abstract
domain the \emph{minimal} amount of information in order to make it
complete, while complete cores act in the opposite direction by
removing the minimal amount of information in order to achieve
completeness. As shown in \cite{jacm}, complete cores always exist,
while complete shells exist under the weak hypothesis that the
concrete semantics $f$ is Scott-continuous. Furthermore, complete
cores and shells enjoy a constructive fixpoint characterization.
While it should be clear that completeness could be achieved by
refining abstract domains, perhaps it is somehow surprising that also
by removing information from an abstract domain one could reach the
completeness property. In this case the abstraction is intended to
remove from an incomplete abstract domain exactly the source of
incompleteness. Let us consider a simple example to illustrate this.
Consider the following abstract domain of signs $\mathit{Sign}^+ \ud
\{\mathbb{Z}, [0,+\infty], [-\infty,0], [0,9],[0]\}$, which
additionally to sign information also represents precisely the
interval $[0,9]$. It turns out that $\mathit{Sign}^+$ is not complete
for integer multiplication: for example, $2\times 3$ is approximated
in $\mathit{Sign}^+$ by $[0,9]$ while the abstract multiplication
$\ok{\alpha_{\mathit{Sign}^+}} (2)
\ok{\times^{\mathit{Sign}^+}}  \ok{\alpha_{\mathit{Sign}^+}} (3)$
gives $[0,+\infty]$. However, $\mathit{Sign} = \{\mathbb{Z}, [0,+\infty],
[-\infty,0],[0]\}$, which is an abstraction of $\mathit{Sign}^+$,
turns out to be complete for multiplication. Even more,
$\mathit{Sign}$ is the most concrete domain which abstracts
$\mathit{Sign}^+$ and is complete for multiplication, namely
$\mathit{Sign}$ is the complete core of $\mathit{Sign}^+$ for
multiplication. Hence, the complete core isolated and removed from
$\mathit{Sign}^+$ the abstract value $[0,9]$, which was the unique
source of incompleteness.

\paragraph*{\textbf{Main Results.}}
We characterize the complete core and shell of the universal state
domain $\wp(\States)$ for all the trace transformers of the
$\mus$-calculus which are sources of incompleteness: negation,
next-time, time-reversal and disjunction.  We also characterize the
structure of transition systems such that the universal state-based
model checking is complete for next-time and time-reversal.  In
particular, disjunction turns out to be the crucial connective. In
fact, the trace-complete shell of the universal state domain for the
disjunction operation is (essentially) the domain of traces itself,
while the trace-complete core is the trivial abstraction of states
carrying no information at all.  Let us point out that one remarkable
feature of our abstract interpretation-based approach lies in the fact
that it is fully constructive, namely we exploit general abstract
interpretation results that always provide complete cores and shells
in fixpoint form.

On the basis of this analysis, we show that for the $\mus$-calculus:
\begin{itemize}
\item[{\rm (1)}] The most abstract refinement of the domain of states
  that induces a model checking which is trace-complete results to be
  the domain of traces itself.
\item[{\rm (2)}] The straightforward abstraction to a non-informative
  singleton is the unique abstraction of the domain of states (and
  hence of the domain of traces) which induces a trace-complete model
  checking.
\item[{\rm (3)}] For each basic temporal/logical operator of the $\mus$-calculus we
  constructively characterize the complete core and shell of the state
  abstraction for traces. These results provide the basis for
  isolating fragments of the $\mus$-calculus which have
  nonstraightforward trace-complete shells and cores of states.
\end{itemize}
These results prove that there is no way to get a complete
approximation of the trace-based semantics by either refining or
approximating the state-based model checking for the entire
$\mus$-calculus, emphasizing the intrinsic limits of the precision of
state-based model checking with respect to the trace-based semantics.
Moreover, since abstract model checking can be
viewed as abstract interpretation of
$\ok{\grass{\cdot}_\mathrm{state}}$ (cf.\ \cite{CC00}), this also
implies that any abstract model checking is intrinsically incomplete
with respect to the trace-semantics of the $\mus$-calculus.

\section{Abstract interpretation and model checking}\label{aib}

\subsection{Notation}
If $X$ is any set then $\ok{\Cl^\cap,\Cl^\cup}:\wp(\wp(X))\ra \wp(\wp(X))$ denote,
respectively, 
the operators that close any subset $Y\in \wp(\wp(X))$ under arbitrary
intersections and unions, e.g.\ $\ok{\Cl^\cap (Y)}\ud  \{ \cap S ~|~ S
\subseteq Y\}$. Note that $X\in \ok{\Cl^\cap(Y)}$ and $\varnothing \in
\ok{\Cl^\cup (Y)}$ because $X=\cap \varnothing$ and $\varnothing = \cup
\varnothing$. If $S\subseteq X$ then $\bneg S$ denotes the
complement of $S$ in $X$. 

A poset $P$ w.r.t.\ a partial ordering $\leq$ is
denoted by $\tuple{P,\leq}$ or $P_\leq$. 
We use the symbol $\sqsubseteq$ to denote pointwise ordering between
functions: if $X$ is any set, $P_\leq$ a poset, and $f,g:X \ra P$ then
$f\sqsubseteq g$ if for all $x\in X$, $f(x)\leq g(x)$.  
If $P$ is a poset and $X\subseteq P$
then $\max (X)\ud \{x\in X~|~\forall y\in X.\: x\leq y \Rightarrow
x=y\}$. 
We denote by $\lfp (f)$ and $\gfp (f)$ (or by $\ok{\lfp^\leq (f)}$
and $\ok{\gfp^\leq (f)}$ to emphasize the partial ordering $\leq$),
respectively, the least and greatest fixpoint, when they exist, of an
operator $f:P\ra P$ on a poset $P_\leq$. It is well known that if 
$\tuple{C,\leq,\vee,\wedge,\top,\bot}$ 
is a complete lattice (actually, a CPO would be enough) and $f:C\ra C$ is
monotone than both $\lfp(f)$ and $\gfp(f)$ exist and
the
following characterizations hold:
$$\lfp(f) = \wedge \{x\in C~|~ f(x) \leq
x\},~~~~~~~\gfp(f) = \vee \{x\in C~|~ x \leq f(x) \}.$$
It also well known that if $f$ is continuous~---~i.e.\ $f$ preserves lub's
of directed subsets or, equivalently, of ascending chains~---~then 
$\lfp(f) =\ok{\vee_{i\in \mathbb{N}} f^{i} (\bot)}$, 
where the sequence $\ok{\{f^i(x)\}_{i\in \mathbb{N}}}$, for any $x\in C$, is 
inductively defined by $f^0 (x) \ud x$ and $f^{i+1} (x)\ud f(f^i (x))$. 
Dually, if $f$ is co-continuous then 
$\gfp(f) =
\ok{\wedge_{i\in \mathbb{N}} f^{i} (\top)}$. A function $f:C\ra C$ is
(finitely) additive when $f$ preserves lub's  of
(finite) arbitrary
subsets of $C$, while co-additivity is dually defined.

\subsection{Abstract interpretation and completeness}\label{aic}
\subsubsection{The lattice of abstract domains}
In standard abstract interpretation \cite{CC77,CC79},
abstract domains can be equivalently specified either by Galois
connections/insertions (GCs/GIs) or by (upper) closure
operators (uco's).  
These two approaches are
equivalent, modulo isomorphic representations of domain's objects. 
The closure operator approach enjoys the advantage of being
independent from the representation of domain's objects:  
in fact, an abstract
domain here is given as a function on the concrete domain of computation. This
feature makes closures appropriate for reasoning on
abstract domains independently from their representation.
Given a complete lattice $C_\leq$, playing the role of concrete
domain, recall that  $\rho:C\ra C$ is a uco when
$\rho$ is monotone, idempotent and extensive (viz.\ $x \leq \rho
(x)$). 
We denote by $\uco(C)$ the set of uco's on $C$.
Let us recall that each $\rho\in \uco(C)$ is
uniquely determined by the set of its fixpoints, which is its image,
i.e.\ $\img(\rho) = \{x\in C ~|~ \rho(x)=x\}$, because $\rho = \lambda
x. \wedge \{ y\in C~|~ y\in \img(\rho), \, x\leq y\}$.  Moreover, a
subset $X\subseteq C$ is the set of fixpoints of some uco on $C$ iff $X$
is meet-closed, i.e.\ $X=\ok{\cM (X)\ud}\{ \wedge Y~|~ Y\subseteq
X\}$ (note~that $\top_C =\wedge \varnothing \in \cM (X)$).    
Note
that when $C=\wp(S)_{\subseteq/\supseteq}$, for some set $S$, then
$\cM=\Cl^\cap\!/\!\Cl^\cup$.   
Often, we will identify closures with their sets of fixpoints.  This
does not give rise to ambiguity, since one can distinguish their use
as functions or sets according to the context. 
It is well known that $\uco(C)$ endowed with the pointwise
ordering $\sqsubseteq$, gives rise to the complete lattice
$\tuple{\uco(C), \sqsubseteq, \sqcup,\sqcap,\lambda x.\top,\id}$.
It turns out that pointwise ordering between uco's corresponds to
superset ordering of the corresponding sets of fixpoints, i.e., $\rho \sqsubseteq \mu$ iff
$\img(\mu)\subseteq \img(\rho)$. 
Let us also recall that for any $\rho \in \uco(C)$ and $X\subseteq C$, 
$\rho (\vee X) = \rho (\vee_{x\in X} \rho(x))$, and for any 
set of closures $\{\rho_i\}_{i\in I}\subseteq \uco(C)$:
$$\sqcup_{i\in I} \rho_i = \cap_{i\in I} \rho_i;~~~~~~~
\sqcap_{i\in I} \rho_i = \cM (\cup_{i\in I} \rho_i );~~~~~~~
\sqcap_{i\in I} \rho_i  = \lambda x. \wedge_{i\in I} \rho_i (x).$$

We denote by $(\alpha,C,A,\gamma)$ a GC/GI of the abstract domain $A$
into the concrete domain $C$ through the abstraction and
concretization maps $\alpha:C\ra A$ and $\gamma:A\ra C$. Thus,
$\alpha$ and $\gamma$ need to form an adjunction between $C$ and $A$:
$\alpha(c)\leq_C a \Lra a \leq_A \gamma(a)$. The map $\alpha$
($\gamma$) is called the left (right) adjoint of $\gamma$
($\alpha$). Let us recall that it is enough to specify either the
abstraction or the concretization map because in any GC the left/right
adjoint map uniquely determines the right/left adjoint map: on the one
hand, any $\alpha:C\ra A$ admits a necessarily unique right adjoint
$\gamma:A \ra C$ defined by $\gamma(a) = \vee_C \{c\in C~|~\alpha(c)
\leq_A a\}$ iff $\alpha$ is additive; on the other hand, any
$\gamma:A\ra C$ admits a necessarily unique left adjoint $\alpha:C \ra
A$ defined by $\alpha(c) = \wedge_A \{a\in A~|~ c \leq_C \gamma(a)\}$
iff $\gamma$ is co-additive.  Recall that a GC is a GI when $\alpha$
is onto or, equivalently, $\gamma$ is 1-1. In abstract interpretation
terms, this means that $A$ does not contain useless abstract values,
namely objects in $A$ which are not abstractions of some concrete
object in $C$. Let us recall that $\rho_A \ud \gamma \circ \alpha$ is
the uco corresponding to the GC $\gdca$ and, conversely, any $\rho \in
\uco(C)$ induces a GI $(\rho,C,\img(\rho),\id)$. Moreover, these two
constructions are one the inverse of each other.  By this equivalence,
throughout the paper, $\tuple{\uco(C),\sqsubseteq}$ will play the role
of the (complete) lattice of abstract domains of the concrete domain
$C$.  The pointwise ordering on $\uco (C)$ corresponds to the standard
order used to compare abstract domains with regard to their precision:
$A_1\sqsubseteq A_2$ in $\uco(C)$ encodes the fact that $A_1$ is more
precise or concrete than $A_2$ or, equivalently, $A_2$ is less precise
or more abstract than $A_1$; in this case, we also say that $A_1$ is a
refinement of $A_2$ and $A_2$ is a simplification or abstraction of
$A_1$.  Lub's and glb's on $\uco(C)$ have therefore the following
reading as operators on abstract domains. Let $\{ A_i\}_{i\in
  I}\subseteq \uco(C)$: (i)~$\sqcup_{i\in I}A_i $ is the most concrete
among the domains which are abstractions of all the $A_i$'s;
(ii)~$\sqcap_{i\in I} A_i $ is the most abstract among the domains
which are more concrete than every $A_i$~---~this domain is also known
as reduced product of all the $A_i$'s.

\subsubsection{Complete abstract domains}\label{cad}
Let $(\alpha,C,A,\gamma)$ be a GI, $f:C\ra C$ be some 
concrete semantic function~---~for simplicity of notation, 
we consider here 1-ary functions~---~and
$f^\sharp:A \ra A$ be a corresponding abstract semantic function. Then,
$\tuple{A,f^\sharp}$ is a sound abstract interpretation, or $f^\sharp$
is a correct approximation of $f$ on $A$, 
when $\ok{\alpha \circ f \sqsubseteq f^\sharp\circ \alpha}$.
The abstract function
$\ok{f^A \ud \alpha \circ f \circ \gamma: A\rightarrow A}$ is called
the best
correct approximation of $f$ in $A$.  Completeness in abstract
interpretation \cite{CC77,jacm} 
corresponds to require the following strengthening of
soundness: 
$\ok{\alpha \circ f = f^\sharp \circ \alpha}$. 
Hence, in addition to soundness, completeness
corresponds to require that no loss of
precision is introduced by the abstract function $\ok{f^\sharp}$
on an approximation $\alpha(c)$ of a 
concrete object $c\in C$ with respect to approximating by $\alpha$
\begin{minipage}{9cm}
the concrete computation $f(c)$.
As a very simple example, let us consider again the abstract domain
$\mathit{Sign}$ representing the sign of an integer variable. 
Let us also consider the binary concrete
operations of
integer addition and multiplication lifted to sets
of integers in $\wp(\bZ)$, e.g., 
$X+Y =\{x+y~|~ x\in X,\, y\in Y\}$.
  Hence, it turns out that the
best correct approximation
$\ok{+^\mathit{Sign}}$ on
$\mathit{Sign}$ of
\end{minipage}
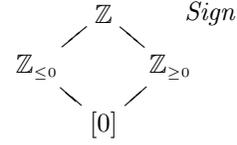
\begin{minipage}{3.6cm}
\begin{center}
    \mbox{      
      \xymatrix@=6pt{
&\mathbb{Z} \ar@{-}[dl]  \ar@{-}[dr] & & *\txt{$\!\!\!\!\!\!\!\mathit{Sign}$}\\
\mathbb{Z}_{\scriptscriptstyle \leq 0} & &
\mathbb{Z}_{\scriptscriptstyle \geq 0} \\
&[0] \ar@{-}[ul]  \ar@{-}[ur]& \\
      }
    }
\end{center}
\end{minipage}
integer addition  is sound but not complete because $\alpha(\{-1\} +
\{1\}) = \alpha(\{0\})=[0] <_{\mathit{Sign}}  \mathbb{Z}  =
\mathbb{Z}_{\scriptscriptstyle \leq 0} \,\, \ok{+^{\mathit{Sign}}}\,\,
\mathbb{Z}_{\scriptscriptstyle \geq 0} = \alpha(\{-1\})
\ok{+^{\mathit{Sign}}} \alpha(\{1\})$. On the other hand, it is
immediate to note that
the best correct approximation of
integer multiplication is instead complete.

Let us recall that 
completeness lifts to least fixpoints, i.e., if
$\tuple{A,f^\sharp}$ is complete then $\alpha (\lfp(f)) =
\lfp(f^\sharp)$. 
Completeness is an abstract domain property because it only depends on
the abstract domain: in fact, it turns out that 
$\tuple{A,f^\sharp}$ is complete iff $\tuple{A,f^A}$ is
complete. Thus, completeness can be equivalently stated as a property
of closures: $A$ is complete iff $\alpha \circ f = f^A \circ
\alpha$ iff $\gamma \circ \alpha \circ f = \gamma \circ \alpha \circ f
\circ \gamma \circ \alpha$. 
Thus, for abstract domains specified as
closure operators, an abstract domain $\rho\in \uco(C)$ is defined to be
complete for $f$ if $\rho \circ f = \rho \circ f \circ \rho$. More in
general, the definition of completeness is extended to
any set $F$ of semantic functions by requiring completeness for each
$f\in F$.  Throughout the paper, we will adopt the following notation:
$\Gamma(C,f)\ud \{\rho \in \uco(C)~|~ \rho \mbox{ is complete for }
f\}$, so that for a set $F$, $\Gamma(C,F)=\cap_{f\in F}\Gamma(C,f)$.
The following property will be useful later on. 
$$\rho \in \Gamma(C,f) \text{~~~iff~~~} \rho \in
\Gamma(C,\{f^n\}_{n\in \mathbb{N}}) \eqno(*)$$
In fact, let us show that by induction on $n\in \mathbb{N}$ that 
if $\rho \in \Gamma(C,f)$ then for any $n\in \mathbb{N}$, $\rho\in\Gamma(C,f^n)$. 
The case $n=0$ amounts to $\rho\in \Gamma(C,\lambda x.x)$ which is
trivially true. For $n+1$ we have that:
$\rho \circ
f^{n+1}=$ (since $\rho\in\Gamma(C,f)$)
$=\rho \circ f \circ \rho  \circ f^n=$ (by inductive
hypothesis)
$=\rho \circ f \circ \rho  \circ f^n \circ \rho=$ (since
$\rho\in\Gamma(C,f)$)
$=\rho \circ f   \circ f^n \circ \rho =
\rho \circ f^{n+1} \circ \rho$.

Let us also recall that, by a well-known result (see, e.g.,
\cite[Theorem~7.1.0.4]{CC79} and \cite[Section~6]{CC00})  
complete abstract domains are ``fixpoint complete'' as well. This
means that if $\rho\in \Gamma(C,f)$, where $f$ is monotone, then
$\lfp(\rho \circ f) = \rho(\lfp(f))$. Moreover, if either $\rho$ does not
contain infinite descending chains or $\rho$ is co-continuous 
then this also holds for greatest fixpoints,
namely $\gfp(\rho \circ f) =
\rho(\gfp(f))$.

\subsubsection{Complete core and shell}\label{ccs}
The fact that  
completeness is an abstract domain property 
opens the question of making an abstract interpretation
complete by minimally extending or, dually, restricting the underlying
abstract domain.  Following \cite{jacm}, given a set of concrete
semantic functions  \mbox{$F \subseteq C\ra C$} and an abstract domain $A\in \uco(C)$, the
\textit{complete shell} (respectively, \textit{core}) of $A$ for
$F$, when it exists, is the most abstract (respectively, concrete) domain
$A^s\in \uco(C)$ (respectively, $A^c\in \uco(C)$) which extends (respectively,
restricts) $A$ and is
complete for $F$.  In other words, the complete shell,
respectively core, of $A$ characterizes the least amount of
information to be added to, respectively removed from, $A$ in order to
get completeness, when this can be done. Complete shell and core of
$A$ for $F$ are denoted, respectively, by $\shell_F(A)$ and
$\core_F(A)$.  
Thus, a complete shell $\shell_F(A)$ exists when $\sqcup \{ A' \in \uco(C)~|~ A'
\sqsubseteq A,\, A' \in \Gamma(C,F)\} \in \Gamma(C,F)$, while a
complete core $\core_F(A)$ exists when $\sqcap \{ A' \in \uco(C)~|~ A
\sqsubseteq A',\, A' \in \Gamma(C,F)\} \in \Gamma(C,F)$.

These problems were 
solved by Giacobazzi et al.~\cite{jacm} who gave a constructive
characterization of complete shells and cores.  
Given a set of functions $F\subseteq C\ra C$, the
abstract domain transformers  $L_F,R_F :\uco(C)\ra \uco(C)$ are defined as
follows: 
\begin{align*}
L_F (\eta) &\ud \{ y \in C~|~
\cup_{f\in F} \max (\{x\in C~|~f(x) \leq y\}) \subseteq
\eta \}\\
R_F (\eta) &\ud \cM (\cup_{f\in F,y\in \eta} \max (\{x\in
C~|~f(x)\leq y\})).
\end{align*}
 
\begin{theorem}[{Giacobazzi et al.~\cite{jacm}}]\label{ft} 
Let $F$ be a set of continuous functions and $\rho \in \uco(C)$. Then,
$\rho \in \Gamma(C,F)$ iff $L_F(\rho) \sqsubseteq \rho$ iff $\rho
\sqsubseteq R_F(\rho)$. Moreover, 
the
complete shell and core of $\rho$ for $F$ exist and are
constructively characterized as follows:
$$\shell_F (\rho) = \sqcap_{i\in \mathbb{N}} R_F^i (\rho),~~~~~~~~
\core_F(\rho) = \sqcup_{i\in \mathbb{N}} L_F^i (\rho).$$ 
\end{theorem}

\noindent
Thus, the complete shell of $\rho$ for $F$ can be obtained by
iteratively adding
to $\rho$ the image of the transformer $R_F$ on the current
domain,  while the complete core can be obtained by iteratively
removing from $\rho$ the elements that are not in the image of the
transformer $L_F$ on the current domain. 

\begin{example}
Let us consider again the abstract domain $\mathit{Sign}^+$
which abstracts $\wp(\mathbb{Z})_\subseteq$ and 
\begin{minipage}{8.5cm}
the square operation
on sets of integers
$sq:\wp(\integer)\ra
  \wp(\integer)$ such that 
$sq(X) = \{x^2~|~x\in X\}$. It turns out that $\mathit{Sign}^+$ is not
complete for $sq$: in fact, $\rho_{\mathit{Sign}^+} (sq
(\rho_{\mathit{Sign}^+} ([0,3]))) = \rho_{\mathit{Sign}^+} (sq
([0,9])) = \mathbb{Z}$, while $\rho_{\mathit{Sign}^+} (sq
([0,3])) = \rho_{\mathit{Sign}^+} (\{0,1,4,9\}) =
[0,9]$. Theorem~\ref{ft} tells us that the abstract element $[0,9]$ 
is a source of incompleteness: in fact, we have that
$\max (\{ X\in \wp(\mathbb{Z})~|~ sq(X) \subseteq [0,9]\}) =[-3,3]$ 
\end{minipage}
\begin{minipage}{4.1cm}
\begin{center}
    \mbox{      
      \xymatrix@=5pt{
&\mathbb{Z} \ar@{-}[dl]  \ar@{-}[dr] & & *\txt{$\!\!\!\!\!\!\!\mathit{Sign}^+$}\\
\mathbb{Z}_{\scriptscriptstyle \leq 0} & &
\mathbb{Z}_{\scriptscriptstyle \geq 0} \\
&& [0,9] \ar@{-}[u] \\
&[0] \ar@{-}[uul]  \ar@{-}[ur]& \\
      }
    }
\end{center}
\end{minipage}
$\not \in \rho_{\mathit{Sign}^+}$ so that $R_{sq} (\rho_{\mathit{Sign}^+}) \not\subseteq
\rho_{\mathit{Sign}^+}$. Moreover, 
$[0,9]$ is the unique source of incompleteness in $\mathit{Sign}^+$ because:

\medskip
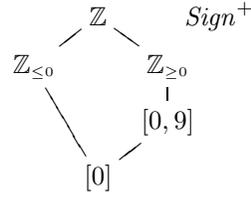
\begin{tabular}{l}
$\max (\{ X\in \wp(\mathbb{Z})~|~ sq(X) \subseteq \mathbb{Z}\}) = \mathbb{Z}
\in \rho_{\mathit{Sign}^+}$ \\[5pt]
$\max (\{ X\in \wp(\mathbb{Z})~|~ sq(X) \subseteq
\mathbb{Z}_{\scriptscriptstyle \leq 0} \}) = \{0\}
\in \rho_{\mathit{Sign}^+}$ \\[5pt]
$\max (\{ X\in \wp(\mathbb{Z})~|~ sq(X) \subseteq
\mathbb{Z}_{\scriptscriptstyle \geq 0} \}) = \mathbb{Z}
\in \rho_{\mathit{Sign}^+}$ \\[5pt]
$\max (\{ X\in \wp(\mathbb{Z})~|~ sq(X) \subseteq
\{0\} \}) = \{0\}\in \rho_{\mathit{Sign}^+}$ 
\end{tabular}

\medskip
\noindent
Thus, by Theorem~\ref{ft}, we have that $\core_{sq}(\mathit{Sign}^+) =
\mathit{Sign}$. 
\ddef
\end{example}

When  $f:C\ra C$ is a mere monotone function in general the complete
shell of an abstract domain for $f$ may not exist, while the complete
core of an abstract domain for $f$ always exists 
even if it cannot be constructively characterized by
Theorem~\ref{ft}. 

\begin{remark} \label{segu}
\rm 
Let $F$ be a set of additive functions. Then, any $F\ni f:C\ra C$ admits a
right adjoint $f^r:C\ra C$ defined by $f^r(y) = \vee \{x\in C~|~f(x)
\leq y\}$. In this case, 
the operators $L_F$ and $R_F$ can be
simplified as follows:
$$L_F (\eta) = \{ y \in C~|~
\{ f^r (y) ~|~ f\in F\} \subseteq \eta \};~~~~~~
R_F (\eta) = \cM (\{ f^r (y) ~|~ y\in \eta, \, f\in F\}).$$
\end{remark}

\subsection{Temporal abstract interpretation}\label{tai}
Let us recall the basic notions and definitions of Cousot and Cousot's
\cite{CC00} temporal abstract interpre\-tation framework (see also
Schimdt's paper \cite{sch01}).
$\mathbb{S}$ is any given, possibly infinite, set of states.  Discrete
time is modeled by the whole set of integers and therefore paths
of
states are time-symmetric, in particular are infinite also in the
past: $\mathbb{P}\ud \mathbb{Z} \ra \mathbb{S}$ is the set of
paths. As usual, an execution path with an initial state $s$ can be encoded
by repeating forever in the past the state $s$. Traces keep track of
the present time, so that $\mathbb{T}\ud \mathbb{Z} \times
\mathbb{P}$ is defined to be the set of traces. We denote by 
$\sigma_i\in \mathbb{S}$ the present state of a trace
$\tuple{i,\sigma}\in \mathbb{T}$.  
The trace-semantics of a temporal formula $\phi$ will be a
temporal model, namely the set of traces making
$\phi$ true.

Temporal models will be generated by transition systems or Kripke
structures, encoding some
reactive system. The transition relation $\sra\, \subseteq
\mathbb{S}\times \mathbb{S}$ is assumed to be (backward and forward) 
total, i.e., $\forall s\in
\mathbb{S}.\exists s' \in \mathbb{S}. \: s \rat s'$ and
$\forall s'\in \mathbb{S}.\exists s \in \mathbb{S}. \: s\rat s' $. 
This is not restrictive, since any transition relation can be
lifted to a total transition relation by adding transitions
$s\rat s$ for any state $s$ which is not reachable (i.e., an initial state) or which cannot
reach any state (i.e., a final state).  The model generated by a transition system
$\tuple{\mathbb{S},\sra}$ is therefore defined as $\mathscr{M}_\sra
\ud \{\tuple{i,\sigma}\in \mathbb{T}~|~ i\in \mathbb{Z},\; \forall
k\in \mathbb{Z}.\: \sigma_k \rat \sigma_{k+1}\}$.
The
pre/post transformers on $\wp(\mathbb{S})$ induced by
$\tuple{\mathbb{S},\sra}$ are defined as usual:
\[
\begin{array}{ll}
- & \pre_\sra (Y) \ud \{ a\in \mathbb{S}~|~\exists b\in Y.\; a \sra
b\};\\
- & \pret_\sra (Y) \ud \bneg (\pre_\sra (\bneg Y)) = 
\{ a\in \mathbb{S} ~|~\forall b\in \mathbb{S}. (a \sra
b \Rightarrow  b\in Y)\}; \\
- & \post_\sra (Y) \ud \{ b\in \mathbb{S}~|~\exists a\in Y.\; a \sra
b\}; \\
- & \postt_\sra (Y) \ud \bneg (\post_\sra (\bneg Y)) = 
\{ b\in \mathbb{S}~|~\forall a\in \mathbb{S}. (a \sra
b \Rightarrow a\in Y)\}.
\end{array}
\]

The forward closure $\mathrm{Fd}:\wp(\mathbb{T})\ra \wp(\mathbb{T})$
is defined as $\mathrm{Fd}(X)\ud \{\tuple{i,\sigma}\in
\mathbb{T}~|~\exists \tuple{i,\tau}\in X. \forall j\geq i. \sigma_j =
\tau_j\}$. Dually, $\mathrm{Bd}(X)\ud \{\tuple{i,\sigma}\in
\mathbb{T}~|~\exists \tuple{i,\tau}\in X. \forall j\leq i. \sigma_j =
\tau_j\}$ is the backward closure of $X\in \wp(\mathbb{T})$.
A set of traces $X$ is forward
(backward) closed when $\mathrm{Fd}(X)=X$ ($\mathrm{Bd}(X)=X$), while
$X$ is state closed  when $X$ is both forward and backward closed. 
Thus, $X$ is forward (backward) closed when the past (future) does not
matter, while $X$ is state closed when the present only matters.

The reversible $\mus$-calculus was introduced by Cousot and
Cousot \cite{CC00} as a past and future time-symmetric
generalization of the $\mu$-calculus,
with a trace-based
semantics.  
Formulae $\phi$ of
the reversible $\mus$-calculus are inductively defined as follows: 
\[
\phi ::= \boldsymbol{\sigma}_S ~|~ \boldsymbol{\pi}_t 
~|~   X 
~|~ \oplus \,\phi  
~|~   \phi^{\curvearrowleft}
~|~   \phi_1 \vee \phi_2 
~|~   \neg \phi 
~|~   \boldsymbol{\mu} X.\phi 
~|~   \boldsymbol{\nu} X.\phi 
~|~   \boldsymbol{\forall} \phi_1 \!:\! \phi_2 
\]
where $S\in\wp (\mathbb{S})$,  
$t\in \wp (\mathbb{S}\times \mathbb{S})$ and 
$X\in \mathbb{X}$, for an infinite set $\mathbb{X}$ of logical
variables. 
The set of $\mus$-calculus formulae is denoted by $\clm$.

Let us give the intuition for the operators of the 
$\mus$-calculus. $\boldsymbol{\sigma}_S$ stands for a state
atomic proposition which holds in traces whose present state is in 
$S$. $\boldsymbol{\pi}_t$ stands for
a transition atomic proposition which holds in traces whose next step is a
transition in $t$.  $\cl$ is
time-reversal that allows to express past/future time modalities from
corresponding future/past time modalities.  
$\oplus$ is the linear temporal next operator (usually denoted by
$\mathrm{X}$). Finally, 
$\boldsymbol{\forall}$ is a generalized universal
quantification with two arguments. 

Let us recall the trace-semantics for the $\mus$-calculus. 
$\mathbb{E}\ud \mathbb{X}\ra \wp(\mathbb{T})$ denotes the set of environments
over $\mathbb{X}$. Given $\xi \in \mathbb{E}$, $X\in\mathbb{X}$ and 
$N\in \wp(\mathbb{T})$, $\xi[X/N]\in\mathbb{E}$ is the environment that acts as
$\xi$ in $\mathbb{X}\smallsetminus\{X\}$ and maps $X$ to $N$. The
$\mus$-calculus semantics $\grass{\cdot}:\clm \ra
\mathbb{E}\ra \wp(\mathbb{T})$ 
is inductively and partially~---~because least or
greatest fixpoints could not exist~---~defined as follows:
$$
\begin{array}{ll}
\grass{\boldsymbol{\sigma}_S}\xi \ud \boldsymbol{\sigma}_{\grassg{S}}&
~~~~~~~~~\grass{\phi_1 \vee \phi_2} \xi \ud \grass{\phi_1}\xi  \cup
\grass{\phi_2}\xi
\\[5pt]
\grass{\boldsymbol{\pi}_t}\xi \ud \boldsymbol{\pi}_{\grassg{t}} &
~~~~~~~~~\grass{\neg \phi} \xi \ud \bneg(\grass{\phi}\xi)
\\[5pt]
\grass{X} \xi \ud \xi (X)&
~~~~~~~~~\grass{\boldsymbol{\mu} X.\phi}\xi \ud  \lfp (\lambda
N\in\wp(\mathbb{T}).\grass{\phi}\xi[X/N]) 
\\[5pt]
\grass{\oplus \,\phi}\xi \ud \boldsymbol{\oplus}\!(\grass{\phi}\xi)  &
~~~~~~~~~\grass{\boldsymbol{\nu} X.\phi}  \xi \ud \gfp (\lambda
N\in\wp(\mathbb{T}). \grass{\phi}\xi[X/N])
\\[5pt]
\grass{\phi^{\curvearrowleft}} \xi \ud 
\cl(\grass{\phi} \xi)&
~~~~~~~~~\grass{\forall \phi_1 \!:\! \phi_2} \xi \ud
\boldsymbol{\forall}(\grass{\phi_1}\xi, \grass{\phi_2}\xi)  
\end{array} $$
where the corresponding temporal
transformers are defined as follows:
\begin{itemize}
\item[--] For any $S\in \wp (\mathbb{S})$, $\boldsymbol{\sigma}_{\grassg{S}} \ud
\{\tuple{i,\sigma}\in\mathbb{T}~|~ \sigma_i \in S\}$ is
the $S$-state model, i.e., the set of traces whose current state
belongs to $S$.
\item[--] For any $t\in \wp (\mathbb{S}\times \mathbb{S})$, $\boldsymbol{\pi}_{\grassg{t}}
\ud \{\tuple{i,\sigma}\in
\mathbb{T}~|~ (\sigma_i,\sigma_{i+1})\in t\}$ is the
$t$-transition model, i.e., the set of traces whose next step is a $t$-transition.
\item[--] $\boplus: \wp(\mathbb{T})\ra \wp(\mathbb{T})$ is the 
next-time or predecessor transformer:\\ 
$\boplus (X)\ud \{\tuple{i-1,\sigma}\in 
\mathbb{T}~|~ \tuple{i,\sigma}\in X\} = 
\{\tuple{i,\sigma}\in 
\mathbb{T}~|~ \tuple{i+1,\sigma}\in X\}$. 
\item[--] $\cl : \wp(\mathbb{T})\ra \wp(\mathbb{T})$ is the
reversal transformer:\\ 
$\cl (X)\ud \{ \tuple{-i,
\lambda k.\sigma_{-k}}\in\mathbb{T}~|~\tuple{i,\sigma}\in X\}$. 
\item[--] $\bneg: \wp(\mathbb{T})\ra \wp(\mathbb{T})$ is the complement:\\ 
$\bneg X\ud\mathbb{T}\smallsetminus X$.
\item[--] Given $s\in \mathbb{S}$, $(\cdot)_{\downarrow s}:\wp(\mathbb{T})\ra
\wp(\mathbb{T})$ is the state projection operator:\\ $X_{\downarrow s}\ud
\{\tuple{i,\sigma}\in X~|~\sigma_i = s\}$.
\item[--] $\fab : \wp(\mathbb{T}) \times \wp(\mathbb{T}) \ra \wp(\mathbb{T})$ is the
universal quantifier:\\ $\fab
(X,Y)\ud\{\tuple{i,\sigma}\in X~|~X_{\downarrow \sigma_i}\subseteq
Y\}$. 
\end{itemize}

\noindent
If $\phi\in \clm$ is a closed formula then the semantics 
$\grass{\phi}\xi$ is independent from the environment $\xi$ and thus we simply write
$\grass{\phi}$. 

The time-reversal operator of the $\mus$-calculus allows to 
express both backward and forward
time modalities. Standard linear and branching 
temporal specification languages like
(past and future) $\mathrm{LTL}$, linear $\mu$-calculus, 
$\mathrm{CTL}^*$, $\mathrm{CTL}$, etc.,
can all be expressed as suitable fragments of the $\mus$-calculus,
since the standard missing operators can be defined as
derived operators. Let us see some examples. 
\\[5pt]
\noindent
~~--~
Previous-time (or successor) $\ominus$: 
$\bominus(X) \ud \cl (\boplus (\cl (X))) =  \{\tuple{i+1,\sigma}\in \mathbb{T}~|~
\tuple{i,\sigma}\in X\} = \{\tuple{i,\sigma}\in \mathbb{T}~|~
\tuple{i-1,\sigma}\in X\}$.
\\[5pt]
\noindent
~~--~
Forward sometime (or finally) $\mathrm{F}$: $\mathbf{F}(X) \ud \lfp (\lambda Y \in
\wp(\mathbb{T}). X \cup \boplus (Y)) = \cup_{n\in \mathbb{N}} \boplus^n (X)$.
\\[5pt]
\noindent
~~--~
Forward globally  $\mathrm{G}$: $\mathbf{G}(X) \ud  \gfp (\lambda Y \in
\wp(\mathbb{T}). X \cap \boplus (Y)) = \cap_{n\in \mathbb{N}} \boplus^n (X)$.
\\[5pt]
\noindent
~~--~
Backward sometime $\mathrm{F}_{\scriptscriptstyle \! -}$:  
$\mathbf{F_{\scriptscriptstyle \! -}}(X) \ud  \cl (\mathbf{F}
(\cl (X))) = \cup_{n\in \mathbb{N}} \bominus^n (X)$. 
\\[5pt]
\noindent
~~--~
Backward globally $\mathrm{G}_{\scriptscriptstyle  -}$:
$\mathbf{G_{\scriptscriptstyle  -}}(X) \ud  \cl (\mathbf{G}(
\cl (X))) = \cap_{n\in \mathbb{N}} \bominus^n (X)$. 

\bigskip
Thus, traces in a  model $\ok{\mathscr{M}_\sra}$ can be defined as
$\ok{\gabba \boldsymbol{\pi}_{\shortrightarrow} \ud 
\mathrm{G}(\boldsymbol{\pi}_{\shortrightarrow}) \wedge
\mathrm{G}_{\scriptscriptstyle -}
(\boldsymbol{\pi}_{\shortrightarrow})}$, so that $\ok{\ma = 
\grasse{\,\gabba \boldsymbol{\pi}_{\shortrightarrow}}}$.
Therefore, standard universal
quantification in $\ok{\mathscr{M}_\sra}$ can be defined as 
$\ok{\forall \phi \ud  \forall \, (\gabba
\boldsymbol{\pi}_{\shortrightarrow})  \!:\! \phi}$,  while
existential quantification is defined by
  $\ok{\exists \phi_1 \!:\! \phi_2 \ud
\neg (\forall
\phi_1 \!:\! \neg \phi_2)}$.

In this framework, the trace-based model checking problem is
as follows.  Let $\mathscr{M}_\sra$ be a model and $\phi\in \clm$ be a  
closed temporal specification.
Then, the universal (existential) 
model checking problem consists in determining whether
$\mathscr{M}_\sra \subseteq
\grass{\phi}$  ($\grass{\phi}
\cap \mathscr{M}_\sra \neq \varnothing$).

\subsection{State-based model checking abstraction} \label{sbmc}

Cousot and Cousot \cite{CC00} show how states can be viewed as an
abstract interpretation of traces through universal or
existential checking abstractions. This abstraction from traces to
states induces a corresponding 
state-based model checking problem which is an approximation of  the concrete
trace-based problem. 

\subsubsection{Universal checking abstraction} \label{uca}
For the universal model checking problem, the right notion of approximation 
is encoded by the superset relation.
In fact, if $\ok{\grasse{\cdot}^\sharp}$ is an approximated
semantics such that  $\ok{\grasse{\phi}^\sharp}\subseteq 
\ok{\grasse{\phi}}$ for any $\phi$,
then the universal abstract verification 
$\mathscr{M}_\sra \subseteq \ok{\grasse{\phi}^\sharp}$ entails 
the concrete one  $\mathscr{M}_\sra \subseteq \ok{\grasse{\phi}}$. Thus,
$\ok{\grasse{\cdot}_1^\sharp} \subseteq \ok{\grasse{\cdot}_2^\sharp}$
means that $\ok{\grasse{\cdot}_2^\sharp}$ is a better approximation
than $\ok{\grasse{\cdot}_1^\sharp}$, so that
sets of traces and states are ordered w.r.t.\ the superset relation: 
$\tuple{\wp(\mathbb{T}),\supseteq}$ and
$\tuple{\wp(\mathbb{S}),\supseteq}$ play, respectively, the role of
concrete and abstract domain.
Let $M\subseteq \mathbb{T}$ be any given model, e.g.\ generated by a
total transition system $\tuple{\mathbb{S},\rightarrow}$.
Traces can be abstracted to states through the universal
quantifier: a set of traces $X \subseteq \mathbb{T}$ is abstracted to the
set of states $s\in \mathbb{S}$ such that any trace in the model $M$
whose present state is $s$ belongs to $X$. Formally, 
the universal checking
abstraction $\alpha^\forall_{M}:\wp(\mathbb{T}) \ra \wp(\mathbb{S})$ 
is defined as follows:
$$\alpha^\forall_{M} (X)\ud \{s\in \mathbb{S}~|~ M_{\downarrow s}
\subseteq X \}.$$ 
Thus, $\alpha^\forall_{M}$
abstracts the trace-semantics  
$\grass{\phi}$ of some 
temporal specification $\phi \in \mus$ 
to the set of (present) states $s$ 
which universally satisfy $\phi$, that is, such that any trace of $M$
with present state $s$ satisfies $\phi$. 
This map is onto (by totality of $\sra$) and 
preserves arbitrary  intersections, 
therefore it induces a a Galois insertion $(\alpha^\forall_M, \wp(\mathbb{T})_\supseteq,
\wp(\mathbb{S})_\supseteq, \gamma^\forall_M)$ where $\gamma^\forall_M$ 
is the right adjoint.  
A set of states $S\in \wp(\mathbb{S})$ is
viewed through the concretization map $\gamma^\forall_M$ 
as an abstract representation for the set of  traces in 
$M$ whose
present state belongs to $S$. Hence, the universal concretization 
$\gamma^\forall_{M}: \wp(\mathbb{S}) \ra \wp(\mathbb{T})$ is defined as
follows: 
$$\gamma^\forall_{M}(S) \ud \{\tuple{i,\sigma}\in
M~|~\sigma_i \in S\}.$$
For our purposes it is helpful to view 
the universal abstraction $(\alpha^\forall_M, \wp(\mathbb{T})_\supseteq,
\wp(\mathbb{S})_\supseteq, \gamma^\forall_M)$ as a 
closure operator in order to make our
analysis independent from specific representations of abstract domains of 
$\wp(\mathbb{T})$.

\begin{definition}\rm
The \emph{universal checking closure} (or simply universal closure) 
relative to a model $M\in
\wp(\mathbb{T})$ is given by $\ok{\rho^\forall_M\ud \gamma^\forall_{M} \circ
\alpha^\forall_{M}}\in \uco(\wp(\mathbb{T})_\supseteq)$. Thus, 
$\ok{\rho^\forall_M} = \lambda X.\{\tuple{i,\sigma}\in M~|~
M_{\downarrow \sigma_i} \subseteq X\}$. \ddef
\end{definition}

Notice that, due to the superset relation, $\rho^\forall_M(X) \subseteq X$. 
The intuition is that $\rho^\forall_M (X)$ throws away from $X$ all
those traces $\tuple{i,\sigma}$ either which are not in $M$~---~these
traces ``do not matter'', since $\alpha^\forall_M (\bneg
M)=\varnothing$~---~or which are in $M$ but whose present state
$\sigma_i$ does not universally satisfy $X$.

Let us observe that, for any $S\in \wp(\mathbb{S})$, 
$\gamma^\forall_M (S) =\cup_{s\in S} M_{\da s}$ and that
the set of fixpoints of $\ok{\rho^\forall_M}$
can be also characterized
as follows: 
$$\rho^\forall_{M} = \{ \gamma_M^\forall (S) 
~|~S\subseteq \mathbb{S}\}\eqno(\ddagger)$$
because $\rho^\forall_M = \{\gamma^\forall_M (\alpha_M^\forall
(T))~|~ T\in \mathbb{T}\} = \{\gamma_M^\forall (S)~|~S\in
\mathbb{S}\}$.

\begin{example}
Consider the two states transition system in Example~\ref{first} 
generating the model $\mtr$. Consider the set
of traces depicted below where the arrows point to the present state:
\begin{align*}
a=~~& \cdots \stackrel{}{1}~\stackrel{}{1}~\stackrel{}{1}~\stackrel{\downarrow}{1}~\stackrel{}{1}~\stackrel{}{1} \cdots \\
b=~~& \cdots \stackrel{}{1}~\stackrel{}{1}~\stackrel{}{1}~
\stackrel{\downarrow}{1}~\stackrel{}{1}~\stackrel{}{1}~\stackrel{}{1}~\stackrel{}{2}
~\stackrel{}{2}~\stackrel{}{2} \cdots
\\
c=~~& \cdots
\stackrel{}{1}~\stackrel{}{1}~\stackrel{}{1}~\stackrel{}{2}~\stackrel{}{2}~\stackrel{}{2}~\stackrel{\downarrow}{2}~\stackrel{}{2}~\stackrel{}{2}~\stackrel{}{2}
\cdots
\\
d=~~& \cdots
\stackrel{}{2}~\stackrel{}{2}~\stackrel{}{2}~\stackrel{\downarrow}{2}~\stackrel{}{2}~\stackrel{}{2}~\stackrel{}{2}~\stackrel{}{1}~\stackrel{}{1}~\stackrel{}{1}
\cdots
\end{align*}
For the set of traces $a$ and $b$ the arrow moves
over~$1$ while in $c$ and $d$ the arrow moves over~$2$.   Let $X=a\cup
b \cup c \cup d$. 
It turns out that $\rho_\mtr^\forall(X)= a \cup b$ because: 
\begin{itemize}
\vspace*{-10pt}
\item[--] the trace $\cdots
\stackrel{}{2}~\stackrel{}{2}~\stackrel{}{2}~\stackrel{\downarrow}{2}~\stackrel{}{2}~\stackrel{}{2}
\cdots$ belongs to $(\mtr)_{\downarrow 2}$ but it does not belong to
$X$, so that $c\cap \rho_\mtr^\forall (X)=\varnothing$;
\item[--] the  traces in $d$ do not belong to $\mtr$, so that $d\cap
\rho_\mtr^\forall (X)=\varnothing$. 
\end{itemize}
As a further example, let consider the formula $\oplus p \in \clm$, where
$p=\boldsymbol{\sigma}_{1}$. We have that $\grass{\oplus p}= \boplus
(\mtr)_{\downarrow 1} = (\mtr)_{\downarrow 1} 
\smallsetminus \{\tuple{i,\sigma}\in (\mtr)_{\downarrow
1}~|~\sigma_{i+1} = 2\}$. Therefore, it turns out that $\ok{\rho_\mtr^\forall} (
\grass{\oplus p}) = \varnothing$. \ddef
\end{example}

In the paper, we will make the following weak assumption on the universal
closure.

\begin{hypothesis}\label{hyppo}
For any universal checking closure
$\rho^\forall_M$, the model $M\in \wp(\mathbb{T})$ is
such that (i)~for any $s\in \mathbb{S}$, $|M_{\downarrow s}| > 1$ and
(ii)~$\boplus (M)=M=\bominus(M)$ and $\boplus (\cl (M))=\,\cl(M)=\bominus
(\cl (M))$. \ddef
\end{hypothesis}

\noindent
Hypothesis~(i) means that for any state $s$, there exist at
least two traces in $M$ with present state $s$, while 
hypothesis~(ii) means 
that $M$ and its reversal $\cl (M)$ are closed for forward and backward time
progresses. These conditions are obviously satisfied by any model 
$\ma$ generated by a total transition system
$\tuple{\mathbb{S},\shortrightarrow}$.

\subsubsection{{Existential checking abstraction}} 
The existential checking abstraction is defined by duality. 
In this case, the relation of approximation is set inclusion, because
$\grasse{\phi}\subseteq \ok{\grasse{\phi}_1^\sharp} \subseteq 
\ok{\grasse{\phi}_2^\sharp}$ and 
$\ok{\grasse{\phi}_1^\sharp} \cap M \neq \varnothing$ imply 
$\ok{\grasse{\phi}_2^\sharp} \cap M \neq \varnothing$.
The Galois insertion 
$(\ok{\alpha^\exists_M}, \wp(\mathbb{T})_\subseteq,
\wp(\mathbb{S})_\subseteq, \ok{\gamma^\exists_M})$ is defined by duality as
follows:
\begin{align*}
\alpha^\exists_{M} (X) & \ud \bneg (\alpha^\forall_M(\bneg (X))) = 
\{s\in \mathbb{S}~|~ M_{\downarrow s}\cap X\neq \varnothing\}\\
\gamma^\exists_M (S) & \ud \bneg (\gamma^\forall_M(\bneg (X))) = 
\{\tuple{i,\sigma}\in \mathbb{T}~|~
(\tuple{i,\sigma}\in M) \,\Ra\, (\sigma_i \in S)\}.
\end{align*}
The intuition is that $\alpha^\exists_{M}$ 
abstracts a given 
trace-semantics  
$\grass{\phi}$ to the set of  states which existentially satisfy $\phi$. 
In this case, the \emph{existential checking closure} relative to a
model $M$ is $\ok{\rho^\exists_M\ud \gamma^\exists_{M} \circ
\alpha^\exists_{M}}\in \uco(\wp(\mathbb{T})_\subseteq)$, that is,
\begin{align*}
\ok{\rho^\exists_M}(X) & = \{\tuple{i,\sigma}\in \mathbb{T}~|~
(\tuple{i,\sigma} \in M) \Rightarrow M_{\downarrow \sigma_i}\cap X\neq
\varnothing\}\\
& = \{\tuple{i,\sigma}\in M~|~M_{\downarrow
\sigma_i}\cap X\neq \varnothing\}\cup \bneg M.
\end{align*} 
Hence, $\rho^\exists_M(X)$ adds to $X$ any trace
which is not in $M$~---~these are meaningless because $\alpha^\exists_M
(\bneg M)=\varnothing$~---~and any trace in $M$ whose present state
existentially satisfies $X$. $\rho^\exists_M$ is dual to
$\rho^\forall_M$ since $\rho^\exists_M = \bneg \circ
\rho^\forall_M \circ \bneg$.
In the following, we will consider the universal abstraction
only, since all the results can be stated and proved by duality 
in the existential case.

\subsubsection{State-based abstract semantics}\label{sbas}
The universal abstraction for some model $M$ (typically $M = \mtr$ for some 
total transition system $\tuple{\mathbb{S},\sra}$)
induces a state-based abstract semantics on $\wp
(\mathbb{S})$ of the $\mus$-calculus which is 
obtained by applying standard abstract interpretation: 
basically, this amounts to abstract any trace
transformer on $\wp(\mathbb{T})$ by the corresponding best correct
approximation on $\wp(\mathbb{S})$ induced by the universal abstraction
$\alpha_M^\forall/\gamma_M^\forall$. For example, the
next-time transformer $\boplus:\wp(\mathbb{T})\ra\wp(\mathbb{T})$ is abstracted
to $\alpha^\forall_{M} \circ \boplus \circ
\gamma^\forall_{M}:\wp(\mathbb{S})\ra\wp(\mathbb{S})$. 

The general scenario is as follows.  $\ok{\mathbb{E}^s\ud
\mathbb{X}\ra \wp (\mathbb{S})}$ is the set of state environments.  The 
state-based abstract semantics 
$\ok{\grass{\cdot}_M^\forall :\clm \ra
\mathbb{E}^s\ra \wp (\mathbb{S})}$ is inductively defined
by replacing each trace transformer
$\mathit{Tr}: \wp(\mathbb{T})\ra \wp(\mathbb{T})$ 
with its corresponding best correct approximation on states
$\ok{\alpha_M^\forall} \circ \mathit{Tr} \circ \ok{\gamma_M^\forall} :
\wp(\mathbb{S}) \ra \wp(\mathbb{S})$. The following lemma
characterizes these best correct approximations. 
  
\begin{lemma} \label{abbrev} \ \\[5pt] \
\begin{tabular}{l}
{\rm (1)~} $\alpha_M^\forall (\boldsymbol{\sigma}_{\grassg{S}}) = S$; \\[5pt] 
{\rm (2)~} $\alpha_{\ma}^\forall (\boldsymbol{\pi}_{\grassg{t}}) = \{s\in
\mathbb{S}~|~ \forall s' \in \mathbb{S}.\: s \sra s' \, \Ra \,
(s,s')\in t\}$; \\[5pt]
{\rm (3)~} $\alpha_M^\forall (\gamma_M^\forall
(S_1) \cup 
\gamma_M^\forall (S_2))  = 
S_1 \cup S_2$;\\[5pt]
{\rm (4)~} $\alpha_M^\forall \circ \bneg \circ \gamma_M^\forall =
\bneg$; \\[5pt]
{\rm (5)~} $\alpha_M^\forall \circ \boplus \circ \gamma_M^\forall =
\pret_\sra$ \\[5pt]
{\rm (6)~} $\alpha_M^\forall (\cl (\gamma_M^\forall (S))) = \{s\in
S~|~M_{\downarrow s} = (\cl \! M)_{\downarrow s}\}$; \\[5pt]
{\rm (7)~} $\alpha_M^\forall (\boldsymbol{\forall}(\gamma_M^\forall (S_1),
\gamma_M^\forall (S_2))) = 
S_1\cap S_2$.  
\end{tabular}
\end{lemma}
\begin{proof}
Point (1) is as follows:
$\alpha_M^\forall (\boldsymbol{\sigma}_{\grassg{S}}) =
\{s\in \mathbb{S}~|~ M_{\downarrow s} \subseteq \{\tuple{i,\sigma}\in
\mathbb{T}~|~ \sigma_i \in S\} \} =
\{s\in \mathbb{S}~|~ (\tuple{i,\sigma} \in M \: \& \: \sigma_i = s) 
\Ra  \sigma_i \in S\}$. Since,  by Hypothesis~\ref{hyppo},
$|M_{\downarrow s}|>1$, we obtain that 
$\{s\in \mathbb{S}~|~ (\tuple{i,\sigma} \in M \: \& \: \sigma_i = s)\, 
\Ra  \, \sigma_i \in S\} = S$.\\
Point~(2) is as follows:
$\alpha_{\ma}^\forall (\boldsymbol{\pi}_{\grassg{t}}) =
\{s\in \mathbb{S}~|~ (\ma)_{\downarrow s} \subseteq \{\tuple{i,\sigma}\in
\mathbb{T}~|~ (\sigma_i,\sigma_{i+1})\in t \} \} =
\{s\in \mathbb{S}~|~ (\tuple{i,\sigma} \in \ma \: \& \: \sigma_i = s) \;
\Ra \; (\sigma_i,\sigma_{i+1})\in t \} =
\{s\in
\mathbb{S}~|~ \forall s' \in \mathbb{S}.\: s \sra s' \, \Ra \,
(s,s')\in t\}$.\\
Point~(3) is as follows:  $\alpha_M^\forall (\gamma_M^\forall
(S_1) \cup \gamma_M^\forall (S_2))  = \alpha_M^\forall (\gamma_M^\forall
(S_1 \cup S_2)) =
S_1 \cup S_2$. \\
Let us consider point~(4) and let us show that 
$\bneg \alpha_M^\forall (\bneg \gamma_M^\forall (S)) =
S$. By \cite[Section~11.7]{CC00}, $\bneg \circ \alpha_M^\forall =
\alpha_M^\exists \circ \bneg$ so that we have that $\bneg \alpha_M^\forall (\bneg
\gamma_M^\forall (S)) = \alpha_M^\exists (\gamma_M^\forall (S)) = 
\{s\in \mathbb{S}~|~M_{\downarrow s} \cap \gamma_M^\forall (S) \neq
\varnothing\}$. By exploiting Hypothesis~\ref{hyppo} which guarantees
that $|M_{\downarrow s}|>1$, it is immediate to prove that 
$\{s\in \mathbb{S}~|~M_{\downarrow s} \cap \gamma_M^\forall (S) \neq
\varnothing\}=S$.
\\
Point~(5) is shown in \cite[Section~11.2]{CC00}.\\
Point~(6) is as follows. By \cite[Section~11.7]{CC00},
$\alpha_M^\forall \circ \cl = \alpha^\forall_{^{\cal \!} M}$. Thus,
$\alpha_M^\forall   (\cl (\gamma_M^\forall (S))) = \{t\in
\mathbb{S}~|~(\cl M)_{\downarrow t} \subseteq  \gamma_M^\forall (S)\} = 
\{t\in \mathbb{S}~|~ \cl (M_{\downarrow t}) \subseteq  \cup_{s\in S} 
M_{\downarrow s}\}$. Since $\cl (M_{\downarrow t}) \subseteq
M_{\downarrow t}$ iff $\cl (M_{\downarrow t}) = M_{\downarrow t}$, we
obtain that $\alpha_M^\forall (\cl (\gamma_M^\forall (S))) = \{s\in
S~|~M_{\downarrow s} = (\cl \! M)_{\downarrow s}\}$. 
\\
Finally, point~(7) is as follows. Observe that $\alpha_M^\forall
(\boldsymbol{\forall}(\gamma_M^\forall (S_1),
\gamma_M^\forall (S_2))) = \{ s\in \mathbb{S}~|~ M_{\downarrow s}
\subseteq \{ \tuple{i,\sigma}\in \gamma_M^\forall (S_1)~|~ 
(\gamma_M^\forall (S_1))_{\downarrow \sigma_i} \subseteq
\gamma_M^\forall (S_2)\} \}$. On the one hand, it is easy to check
that $S_1 \cap S_2 \subseteq \alpha_M^\forall
(\boldsymbol{\forall}(\gamma_M^\forall (S_1),
\gamma_M^\forall (S_2)))$. The reverse inclusion follows easily by
noting that Hypothesis~\ref{hyppo} ensures that  for any $s\in \mathbb{S}$ there
exists some $\tuple{i,\sigma}\in M_{\downarrow s}$. 
\end{proof}

By the above lemma, 
the abstract semantics $\ok{\grass{\cdot}_{\ma}^\forall :\clm \ra
\mathbb{E}^s\ra \wp (\mathbb{S})}$ is inductively defined as follows:

\medskip
\begin{tabular}{l}
$\grass{\boldsymbol{\sigma}_S}_{\ma}^\forall \chi = S$\\[5pt]
$\grass{\boldsymbol{\pi}_t}_{\ma}^\forall \chi = \{   s\in
\mathbb{S}~|~ \forall s' \in \mathbb{S}.\: s \sra s' \, \Ra \,
(s,s')\in t\}$\\[5pt]
$\grass{X}_{\ma}^\forall \chi = \chi (X)$\\[5pt]
$\grass{\phi_1 \vee \phi_2}_{\ma}^\forall \chi = \grass{\phi_1}_{\ma}^\forall
\chi \cup \grass{\phi_2}_{\ma}^\forall \chi$\\[5pt]
$\grass{\neg \phi}_{\ma}^\forall \chi =
\bneg \grass{\phi}_{\ma}^\forall \chi$\\[5pt]
\end{tabular}

\begin{tabular}{l}
$\grass{\oplus \,\phi}_{\ma}^\forall \chi =
\pret_\sra (\grass{\phi}_{\ma}^\forall \chi)$\\[5pt]
$\grass{\phi^{\curvearrowleft}}_{\ma}^\forall 
\chi = \alpha_{\ma}^\forall (\cl (\gamma_{\ma}^\forall (
\grass{\phi}_{\ma}^\forall 
\chi)))$\\[5pt]
$\grass{\boldsymbol{\mu} X.\phi}_{\ma}^\forall \chi =  \lfp (\lambda
S\in\wp(\mathbb{S}).\grass{\phi}_{\ma}^\forall \chi[X/S]) $\\[5pt]
$\grass{\boldsymbol{\nu} X.\phi}_{\ma}^\forall  \chi = \gfp (\lambda
S\in\wp(\mathbb{S}). \grass{\phi}_{\ma}^\forall \chi[X/S])$\\[5pt]
$\grass{\forall \phi_1 : \phi_2}_{\ma}^\forall \chi =
\grass{\phi_1}_{\ma}^\forall\chi \cap
\grass{\phi_2}_{\ma}^\forall\chi$
\end{tabular}

\medskip
Thus, for any linear formula $\phi$, namely a formula $\phi$ with no
quantifier, $\ok{\grass{\phi}_{\ma}^\forall}$ provides the
state-semantics of the state formula $\ok{\phi^\forall}$ which is obtained
from $\phi$ by preceding each linear temporal operator, i.e.\ next-time
$\oplus$ and time-reversal $\cl$, occurring in $\phi$ by the universal
path quantifier $\forall$.

The universal abstraction $\ok{\alpha_M^\forall}$ is extended pointwise
to environments $\ok{\dot{\alpha}_M^\forall
:\mathbb{E}\ra\mathbb{E}^s}$ as follows:
$\ok{\dot{\alpha}_M^\forall(\xi)\ud \lambda
X\in\mathbb{X}.\alpha_M^\forall (\xi(X))}$.  The correctness of the
state-based semantics $\ok{\grass{\cdot}_{\ma}^\forall}$ is a
consequence of its abstract interpretation-based definition:
$$\text{For any~}\phi\in \clm \text{ ~and~ } \xi\in \mathbb{E},\;
\ok{\alpha_M^\forall (\grass{\phi}\xi) \supseteq
\grass{\phi}_M^\forall \dot{\alpha}_M^\forall (\xi)}.$$
This means that given any state $s\in \grass{\phi}_M^\forall
\dot{\alpha}_M^\forall (\xi)$, it turns out that any trace
$\tuple{i,\sigma}$ in $M$ whose
present state is $s$ satisfies $\phi$. 
Following the terminology by
Kupferman and Vardi~\cite{kv98,var98}, when 
$\ok{\alpha_M^\forall (\grass{\phi}\xi) =
\grass{\phi}_M^\forall \dot{\alpha}_M^\forall (\xi)}$ holds for some 
$\phi \in \clm$, the formula $\phi$ is called \emph{branchable}.  
In general, completeness does not hold for all the formulae of the $\mus$-calculus, 
i.e.\ the above containment may be strict, as shown in the
Introduction. This intuitively means that  
universal model checking of
linear formulae cannot be reduced with no loss of precision to universal model checking  
on states through the universal quantifier abstraction.
Consequently, it turns out that the universal abstraction is
incomplete for some trace operators of the $\mus$-calculus. 
Cousot and Cousot \cite[Section 11]{CC00} identified the sources of this
incompleteness, namely those
operators $\mathit{Op}$ of the 
$\mus$-calculus such that $\ok{\rho_M^\forall}$ is incomplete for
$\mathit{Op}$: next-time, disjunction, negation and 
time-reversal. Incompleteness of $\ok{\rho_M^\forall}$ w.r.t.\
time-reversal and negation is
not explicitly mentioned in \cite{CC00} and is 
shown by the following example. 

\begin{example}\label{nct}
Consider the two states transition system 
in Example~\ref{first}.  Let $X\ud 
\{ \tuple{i,\sigma}\in \mathbb{T} ~|~\forall k\geq i. \sigma_k = 1\}$,
so that  $\cl (X) = 
\{ \tuple{i,\sigma} ~|~\forall k\leq i. \sigma_k = 1\}$. 
Since  $\ok{(\mtr)_{\downarrow 1}} \not\subseteq X$ and
$\ok{(\mtr)_{\downarrow 2}} \not\subseteq X$, we have that
$\ok{\rho_\mtr^\forall (X)}=\varnothing$  and 
therefore $\ok{\rho_\mtr^\forall (\cl (\rho_\mtr^\forall
(X)))}=\varnothing$. Instead, it turns out that $\ok{\rho_\mtr^\forall (\cl
(X))}= 
(\mtr)_{\downarrow 1}$. This means that $\ok{\rho_\mtr^\forall}$ is not
complete for $\cl$.\\
As far as negation is concerned, consider any 
$\tuple{i,\sigma}\in \ok{(\mtr)_{\downarrow 1}}$ (e.g., $\tuple{0,\lambda
k\in \mathbb{Z}.1}$) and $\tuple{j,\tau}\in \ok{(\mtr)_{\downarrow
2}}$ (e.g., $\tuple{0,\lambda k\in \mathbb{Z}.2}$), and 
let $X\ud \bneg \{\tuple{i,\sigma}, \tuple{j,\tau}\}$.  
Then, it turns out that $\ok{\rho_\mtr^\forall (\bneg X)} =
\ok{\rho_\mtr^\forall (\{\tuple{i,\sigma}, \tuple{j,\tau}\} )}=
\varnothing$,
while $\ok{\rho_\mtr^\forall (\bneg \rho_\mtr^\forall (X))}=
\ok{\rho_\mtr^\forall (\bneg \varnothing)}=   \ok{\rho_\mtr^\forall
(\mathbb{T})}= \ok{\mtr}$, so that completeness does not hold. 
\ddef 
\end{example}

Cousot and Cousot \cite{CC00} provide some conditions on the
incomplete trace operators that ensure completeness of
$\ok{\rho_M^\forall}$. As far as next-time is concerned,
Cousot and Cousot show that completeness of $\rfm$ for
$\ok{\boplus}$ holds when the linear operator $\oplus$ is restricted to
forward closed (i.e.\ future-time) formulae, namely formulae of the
$\mus$-calculus without time-reversal.  On the other hand, when
disjunction is restricted to have at least one state formula, i.e.\ a
universally quantified formula, it turns out that $\rho_M^\forall$ is
complete. These sufficient conditions allow to identify some complete
fragments of the $\mus$-calculus. This is the case, for example, of
the \mbox{$\ok{\mu_+^\forall}$-calculus} considered by Cousot and Cousot in
\cite[Section~13]{CC00}, where time-reversal is disallowed and 
disjunction is restricted to at least
one state formulae. 

Completeness of $\ok{\rho_M^\forall}$ 
is related to Maidl's \cite{maidl00} characterization of 
the maximum common fragment $\ltldet$ of $\ltl$ and $\actl$, which is
defined as follows:
\[
\begin{array}{rcl}
\ltldet \ni \phi & ::= & \boldsymbol{\sigma}_S~|~\neg
\boldsymbol{\sigma}_S~|~\phi_1\wedge\phi_2 ~|~ 
(\boldsymbol{\sigma}_S\wedge \phi_1)\vee(\neg \boldsymbol{\sigma}_S \wedge
\phi_2) ~|~\\[5pt]
& & \oplus \phi~|~\U(\boldsymbol{\sigma}_S \wedge \phi_1,
\neg \boldsymbol{\sigma}_S \wedge \phi_2)
~|~\W(\boldsymbol{\sigma}_S\wedge \phi_1,\neg \boldsymbol{\sigma}_S \wedge \phi_2)
\end{array}
\]
where $\U$ and $\W$ denote, respectively, standard until and
weak-until (i.e., $\W(\phi_1,\phi_2) = \mathrm{G} \phi_1 \vee
\U(\phi_1,\phi_2)$)
operators. Obviously, $\ltldet$ is a fragment of the $\mus$-calculus. 
Maidl~\cite{maidl00} shows that $\ltldet = \ltl \cap \actl$, namely
for any $\phi \in \ltl$, there exists some $\psi \in \actl$ such that 
$\alpha_M^\forall (\grasse{\phi}) = \grasse{\psi}$ iff there exists
some $\zeta \in \ltldet$ such that $\grasse{\phi} =
\grasse{\zeta}$.  

Ranzato and Tapparo \cite{rt05bis} show that the universal abstraction is
complete for all the formulae of $\ltldet$, namely 
for any $\phi\in \ltldet$, $\ok{\alpha_M^\forall (\grass{\phi})=
\grass{\phi}_M^\forall}$. Let $\ltl_\forall = \{ \phi \in
\ltl~|~ \ok{\alpha_M^\forall} (\grass{\phi})=
\ok{\grass{\phi}_M^\forall}\}$ denote the set of branchable $\ltl$
formulae. Thus, we have that $\ltldet \subseteq
\ltl_\forall$. Furthermore, the following converse holds: any
branchable $\ltl$ formula is equivalent to some formula in $\ltldet$.
In fact, if $\phi \in \ltl$ is branchable then, by Maidl's~\cite{maidl00}
Corollary 1, there exists some $\psi\in \ltldet$ such that
$\grass{\phi}=\grass{\psi}$. As a consequence, we obtain
the following  characterization of 
branchability for $\ltl$ formulae. 
\begin{theorem}
Let $\phi \in \ltl$. Then, there exists $\zeta \in \ltl_\forall$ such
that $\grasse{\phi}=\grasse{\zeta}$ if and only if 
there exists $\psi \in \ltldet$ such that 
$\grasse{\phi} = \grasse{\psi}$.
\end{theorem} 
Thus, $\ltldet$ also provides a synctatic characterization for 
the set of  branchable $\ltl$ formulae.

\section{Complete cores and shells for temporal connectives}

In the following, we will characterize the complete cores and
shells of the universal abstraction $\rfm$ for the
following trace operators which are
sources of incompleteness:
next-time, disjunction and time-reversal. 
These complete cores
and shells do exist because $\boplus$, $\cup$ and $\cl$ are trivially
continuous functions on the concrete domain
$\wp(\mathbb{T})_\supseteq$ so that we can exploit Theorem~\ref{ft} in
order to characterize them.  As recalled in Section~\ref{ccs}, 
complete shells may not exist and we show that this is indeed the case of
negation. Let us observe that Theorem~\ref{ft} cannot be applied in
this case 
because negation is not continuous on
$\wp(\mathbb{T})_\supseteq$. On the other hand, the complete core for negation 
does exist. 

One remarkable feature of 
our approach lies in the fact  that it is fully constructive, namely
Theorem~\ref{ft} always provides 
complete cores and shells in fixpoint form 
so that we do not need to conjecture 
some abstract domain and successively to prove that
it is indeed a complete core or shell.

\subsection{Negation}\label{secneg}

\begin{theorem}\label{noneg}
The complete shell of $\ok{\rho_M^\forall}$ for $\bneg$ does not
exist. 
\end{theorem}
\begin{proof}
Let us consider the simplest transition system $\tuple{\{\bullet\},\{\bullet\!\sra\!
\bullet\}}$ consisting of a single state $\bullet$ and of a single transition $\bullet
\!\sra\! \bullet$. The only possible path is $\lambda n\in\mathbb{Z}.\bullet$ so that the 
model $M$ generated by this transition system coincides with the set
of traces, namely
$M=\{\tuple{i,\lambda n.\bullet}~|~ i\in \mathbb{Z}\}$. Thus, any set of
traces can be simply represented by the corresponding
set of present times, namely by a corresponding set of integers, so that the
concrete domain $\wp(\mathbb{T})_\supseteq$ can be represented by
$\wp(\mathbb{Z})_\supseteq$ and in particular $M=\mathbb{Z}$. We also have that
$\ok{\rho_M^\forall} = \{\varnothing, \mathbb{Z}\}$. \\
Let $\mathbb{Z}_{\mathrm{ev}}$ and 
$\mathbb{Z}_{\mathrm{od}}$ denote, respectively, the set of even and
odd intergers and consider the following two closures: for any 
$X\in \wp(\mathbb{Z})$, 
\[
\begin{array}{cc}
\rho_{\mathrm{ev}} (X) = \left\{ \begin{array}{ll}
\mathbb{Z} & \text{if $X=\mathbb{Z}$}\\ 
X \cap  \mathbb{Z}_{\mathrm{ev}} & \text{otherwise} 
\end{array} \right.
~~~~&~~~~
\rho_{\mathrm{od}} (X) = \left\{ \begin{array}{ll}
\mathbb{Z} & \text{if $X=\mathbb{Z}$}\\ 
X \cap \mathbb{Z}_{\mathrm{od}} & \text{otherwise} 
\end{array} \right.
\end{array}
\]
Let us note that $\rho_{\mathrm{ev}}, \rho_{\mathrm{od}} \in
\uco(\wp(\mathbb{Z})_\supseteq)$, because their images are closed under
arbitrary unions, and that $\rho_{\mathrm{ev}}, \rho_{\mathrm{od}} \sqsubseteq \rfm$. 
Let us show that $\rho_{\mathrm{ev}}$ is complete for $\bneg$
(the case of $\rho_{\mathrm{od}}$ is analogous). If
$X\in \{\mathbb{Z},\varnothing\}$ then $\rho_{\mathrm{ev}}(\bneg X) = 
\rho_{\mathrm{ev}} (\bneg \rho_{\mathrm{ev}} (X))$ trivially holds. If
$X\in \wp( \mathbb{Z})$ and $X\not\in \{\mathbb{Z},\varnothing\}$ then
\[
\begin{array}{l}
\rho_{\mathrm{ev}} (\bneg \rho_{\mathrm{ev}} (X)) =
\rho_{\mathrm{ev}} (\bneg (\mathbb{Z}_\mathrm{ev}\cap X )) =
\rho_{\mathrm{ev}} (\mathbb{Z}_\mathrm{od}\cup \bneg X )) =\\[5pt]
\mathbb{Z}_\mathrm{ev} \cap (\mathbb{Z}_\mathrm{od}\cup \bneg X ) =
\mathbb{Z}_\mathrm{ev} \cap \bneg X  =
\rho_{\mathrm{ev}} (\bneg X).
\end{array}
\]
If $\shell_{\bneg} (\rfm)$ would
exist then we would have that $\rho_{\mathrm{ev}},\rho_{\mathrm{od}} \sqsubseteq
\shell_{\bneg} (\rfm)$, so that $ \rho_{\mathrm{ev}}\sqcup \rho_{\mathrm{od}}  \sqsubseteq
\shell_{\bneg} (\rfm)$. But $\rho_{\mathrm{ev}}\sqcup \rho_{\mathrm{od}}=
\rfm$, so that we would have that    
$\shell_{\bneg} (\rfm) = \rfm$ which is a contradiction because
$\rfm$ is not complete for $\bneg$. 
\end{proof}

Negation is antimonotone, however this is not why the corresponding complete shell
does not exist. In fact, as a further remarkable example, 
we show that 
this is also the case for the ``sometime'' operator $\mathrm{F}$,
which is instead monotone.  
\begin{theorem}\label{nof}
The complete shell of $\ok{\rho_M^\forall}$ for $\mathbf{F}$ does not
exist. 
\end{theorem}
\begin{proof}
Let us consider again the transition system $\tuple{\{\bullet\},\{\bullet\!\sra\!
\bullet\}}$ used in the proof of Theorem~\ref{noneg} so that the
concrete domain $\wp(\mathbb{T})_\supseteq$ can be represented by
$\wp(\mathbb{Z})_\supseteq$ and in particular $M=\mathbb{Z}$. We also have that
$\ok{\rho_M^\forall} = \{\varnothing, \mathbb{Z}\}$, namely $\ok{\rho_M^\forall
(\mathbb{Z})}=\mathbb{Z}$, while if $X\subsetneq \mathbb{Z}$ then
$\ok{\rho_M^\forall(X)}=\varnothing$. Let us observe that for any
$k\in \mathbb{Z}$, $\mathbf{F}([k,+\infty))=\mathbb{Z}$, because for any $i\in
\mathbb{Z}$ there exists some $m\geq i$ and $m\in [k,+\infty)$. 

\noindent
It is now simple to observe that
$\ok{\rho_M^\forall}$ is not complete for $\mathbf{F}$. In fact,  for any $k\in
\mathbb{Z}$, we have that $\rfm (\mathbf{F}([k,+\infty))) = \rfm
(\mathbb{Z})=\mathbb{Z}$, while $\rfm (\mathbf{F}(\rfm([k,+\infty)))) = \rfm
(\mathbf{F}(\varnothing)) = \rfm(\varnothing)=\varnothing$. It is also
easy to note that $\mathbf{F}$ is not continuous on
$\wp(\mathbb{T})_\supseteq$: $\bigcap_{k\in \mathbb{Z}}
\mathbf{F}([k,+\infty)) = \mathbb{Z}$, whereas $\mathbf{F} (\bigcap_{k\in
\mathbb{Z}}[k,+\infty)) = \mathbf{F}(\varnothing)=\varnothing$. Hence,
noncontinuity of $\mathbf{F}$ is consistent with Theorem~\ref{ft}. 

\noindent
Let us now consider the following family of closures: for any $k\in
\mathbb{Z}$ and $X\in \wp(\mathbb{Z})$, 
\[
\rho_k (X) = \left\{ \begin{array}{ll}
\mathbb{Z} & \text{if $X=\mathbb{Z}$}\\ 
X \cap [k,+\infty) & \text{otherwise} 
\end{array} \right.
\]
Let us note that $\rho_k \in
\uco(\wp(\mathbb{Z})_\supseteq)$, because $\img(\rho_k)=\{\mathbb{Z}\} \cup \{X
\in \wp(\mathbb{Z})~|~ X\subseteq [k,+\infty)\}$ is closed under
arbitrary unions, and that $\rho_k \sqsubseteq \rfm$. 
Let us show that $\rho_k$ is complete for $\mathbf{F}$. 
Let $X\in
\wp(\mathbb{Z})$. If $X=\mathbb{Z}$ then $\rho_k 
(\mathbf{F}(X)) = \rho_k(\mathbf{F}(\rho_k(X)))$ trivially holds
because $X=\mathbb{Z}\in \rho_k$. Thus, consider $X\subsetneq
\mathbb{Z}$. We distinguish the following two cases.\\
Case~(i).   Assume that for any $j\in \mathbb{Z}$, $X\cap
[j,+\infty)\neq \varnothing$. Then, we have that $\mathbf{F}(X)=\mathbb{Z}$
because, by hypothesis on $X$, 
for any $i\in\mathbb{Z}$ there exists some $k\in X$ such that
$i\leq k$. Moreover, $\mathbf{F}(\rho_k (X))= \mathbf{F}(X\cap
[k,+\infty))= \mathbb{Z}$ because for any $i\in \mathbb{Z}$, $X\cap
[k,+\infty) \cap [i,+\infty) \neq \varnothing$. Thus, in this case, 
$ \mathbf{F}(X) = \mathbf{F}(\rho_k(X))$, so that 
$\rho_k 
(\mathbf{F}(X)) = \rho_k(\mathbf{F}(\rho_k(X)))=\mathbb{Z}$.\\
Case~(ii). 
 On the other hand,
assume that there exists some $i\in \mathbb{Z}$ such that $X\cap
[i,+\infty)=\varnothing$. Therefore, $\max(X)=n\in \mathbb{Z}$ so that
$\mathbf{F} (X) =(-\infty,n]$. Let us distinguish two cases: $n < k$
and $n\geq k$. If $n<k$ then $\rho_k (\mathbf{F}(X)) = (-\infty,n]
\cap [k,+\infty) = \varnothing$,  $\rho_k (X) = X \cap
[k,+\infty)=\varnothing$, so that $\rho_k (\mathbf{F}(\rho_k
(X)))=\varnothing$. If, instead, $n\geq k$ then $\rho_k (\mathbf{F}(X))
= (-\infty,n] \cap [k,+\infty) = [k,n]$, $\rho_k (X) =X \cap
[k,+\infty)$ so that $\max(\rho_k (X))=n$ and this implies 
$\mathbf{F}(\rho_k(X)) = (-\infty,n]$, from which $\rho_k (\mathbf{F}
(\rho_k (X))) = (-\infty,n]\cap [k,+\infty) = [k,n]$. \\
Hence, summing
up, we have shown that for any $k\in \mathbb{Z}$ and $X\in \wp(\mathbb{Z})$, 
$\rho_k 
(\mathbf{F}(X)) = \rho_k(\mathbf{F}(\rho_k(X)))$, i.e.\ any $\rho_k$
is complete for $\mathbf{F}$. If $\shell_{\mathbf{F}} (\rfm)$ would
exist then we would have that for any $k$, $\rho_k \sqsubseteq
\shell_{\mathbf{F}} (\rfm)$, so that $\sqcup_{k\in \mathbb{Z}} \rho_k  \sqsubseteq
\shell_{\mathbf{F}} (\rfm)$. But $\img(\sqcup_{k\in \mathbb{Z}} \rho_k) =
\bigcap_{k\in \mathbb{Z}} \img(\rho_k) =
\{\varnothing,\mathbb{Z}\}=\img(\rfm)$, so that we would have that    
$\shell_{\mathbf{F}} (\rfm) = \rfm$ which is a contradiction because
$\rfm$ is not complete for $\mathbf{F}$. 
\end{proof}

The above proof also shows that  $\mathbf{F}$ is not continuous on
$\wp(\mathbb{T})_\supseteq$, so that 
noncontinuity of $\mathbf{F}$ is consistent with Theorem~\ref{ft}. 

Although negation is not monotone, it turns out that 
the core of $\rfm$ for $\bneg$ exists even if we cannot
exploit Theorem~\ref{ft} in order to obtain a constructive
characterization of it.  
 This core results to be  the greatest totally uninformative closure.

\begin{theorem}\label{coreneg}
$\core_{\bneg}(\rfm) = \lambda X. \varnothing$. 
\end{theorem}
\begin{proof}
Let $\eta \in \uco(\wp(\mathbb{T})_\supseteq)$ such that $\rfm
\sqsubseteq \eta$, so that, for any $X$, $\rfm(X)\supseteq \eta(X)$. 
By Hypothesis~\ref{hyppo},
for any $s\in \mathbb{S}$, we consider some $\tuple{i,\sigma_s}\in 
M_{\downarrow s}$, so that $|M_{\downarrow s}\smallsetminus
\{\tuple{i,\sigma_s}\}| \geq 1$. Consider $Y \ud \{
\tuple{i,\sigma_s}\in \mathbb{T}~|~ s\in \mathbb{S}\}$. Then, we have
that $\eta(\bneg Y)\subseteq \rfm(\bneg Y) = \varnothing$, so that
$\eta(\bneg Y)=\varnothing$. On the other hand, $\eta(Y)\subseteq
\rfm(Y)=\varnothing$, so that $\eta(Y)=\varnothing$ and in turn 
$\eta(\bneg \eta (Y)) = \eta(\bneg \varnothing) = \eta
(\mathbb{T})$. Thus, if $\eta$ is complete for $\bneg$ then
$\eta(\mathbb{T})=\varnothing$ so that for any $X\subseteq
\mathbb{T}$, $\eta(X)\subseteq \eta(\mathbb{T})=\varnothing$. Hence,
$\lambda X.\varnothing$ is the unique closure which is greater than
$\rfm$ and complete for $\bneg$, i.e., $\core_{\bneg}(\rfm) = \lambda
X. \varnothing$.  
\end{proof}


\subsection{Next-time}

Let us first show the following easy properties of the predecessor and
successor trace operators. 

\begin{lemma}\label{preli} \ \\ \
{\rm (1)} $\boplus:\wp(\mathbb{T})\ra \wp(\mathbb{T})$ and
$\bominus:\wp(\mathbb{T})\ra\wp(\mathbb{T})$ preserve arbitrary unions and
intersections, and $\boplus^{-1}=\bominus$ and $\bominus^{-1}=\boplus$. \\ 
Let $\rho \in \uco(\wp(\mathbb{T})_{\supseteq})$.  Then,\\ 
{\rm (2)} $\rho
\in \Gamma(\wp(\mathbb{T})_{\supseteq},\boplus)$ iff for all
$n\in \mathbb{N}$ and $X\in \wp(\mathbb{T})$, $\bominus^n (\rho (X))=\rho
(\bominus^n (\rho (X)))$; \\
{\rm (3)} $\rho
\in \Gamma(\wp(\mathbb{T})_{\supseteq},\bominus)$ iff for all
$n\in \mathbb{N}$ and $X\in \wp(\mathbb{T})$, $\boplus^n (\rho (X))=\rho
(\boplus^n (\rho (X)))$.
\end{lemma}
\begin{proof} 
(1): Clear. \\
(2) and (3): Let us check that $\rho
\in \Gamma(\wp(\mathbb{T})_{\supseteq},\boplus)$ iff for all
$n\in \mathbb{N}$ and $X\in \wp(\mathbb{T})$, $\bominus^n (\rho (X))=\rho
(\bominus^n (\rho (X)))$ (the remaining proof is analogous). 
Because, by~(1), 
$\boplus$ is additive on $\wp(\mathbb{T})_\supseteq$, by
Theorem~\ref{ft} and Remark~\ref{segu}, 
we have that $\rho
\in \Gamma(\wp(\mathbb{T})_{\supseteq},\boplus)$ iff 
$\{\cap \{X\in\wp(\mathbb{T})~|~\boplus (X)\supseteq Y\}\}_{Y \in \rho}
\subseteq \rho$. By~(1), $\boplus (X)\supseteq Y$ iff
$X\supseteq \bominus (Y)$, and therefore 
$\rho
\in \Gamma(\wp(\mathbb{T})_{\supseteq},\boplus)$ iff 
$\{\bominus (Y)~|~ Y\in \rho\}
\subseteq \rho$, and therefore, iff $\{\bominus (\rho(X))~|~ X\in\wp(\mathbb{T})\}
\subseteq \rho$. Analogously, we get that, for any $n\in\mathbb{N}$, $\rho
\in \Gamma(\wp(\mathbb{T})_{\supseteq},\boplus^n)$ iff 
$\{\bominus^n (\rho(X))~|~ X\in\wp(\mathbb{T})\}
\subseteq \rho$. Thus, property~$(*)$ in Section~\ref{cad} closes the proof.   
\qedhere \end{proof}

Let us recall from \cite{CC00} that $\rfm$ is complete for $\oplus$
when $\oplus$ is restricted to
forward closed set of traces, namely if $X\in \wp(\mathbb{T})$ is such
that $X=\mathrm{Fd}(X)$ then $\rfm(\oplus (X)) = \rfm(\oplus (\rfm
(X)))$. This implies that for forward or state closed specification
languages, namely languages with no past-time modality like $\ltl$ and
$\CTLstar$, the universal abstraction is already complete for
the next-time trace transformer. The situation changes in the general
case of the $\mus$-calculus, where $\rfm$ is incomplete for next-time.

\subsubsection{Complete core} 

By exploiting the constructive method provided by Theorem~\ref{ft}, 
the set of fixpoints of the 
complete core $\core_\boplus (\rho^\forall_{M})$ is
first characterized as follows.

\begin{theorem}\label{teosu}
The set of fixpoints of $\core_\boplus (\rho^\forall_{M})$ is
$\ok{\{Y \in \wp(\mathbb{T})~|~ \forall k\in \mathbb{N}.\, \bominus^k Y =
\rho^\forall_{M} (\bominus^k Y)\}}$.
\end{theorem}
\begin{proof}
By Theorem~\ref{ft} and Remark~\ref{segu}, 
$\ok{\core_\boplus (\rho^\forall_{M})} = \ok{\sqcup_{i\in \mathbb{N}} L_F^i
(\rfm)}$. Thus, 
$Y \in 
\ok{\core_\boplus (\rho^\forall_{M})}$ $\Lra$ $\forall i\in \mathbb{N}. Y\in
\ok{L_\boplus^i (\rho^\forall_M)}$.
Moreover, by Lemma~\ref{preli}, we have that $L_\boplus (\eta) =
\{ Y \in \wp(\mathbb{T})~|~
\cap \{X\in \wp(\mathbb{T})~|~ X \supseteq \bominus Y\} \in \eta \}
= \{ Y \in \wp(\mathbb{T})~|~
\bominus Y \in \eta \}  
= \{ Y \in \wp(\mathbb{T})~|~
\bominus Y  =\eta(\bominus Y) \}$, and therefore, for any $i\in \mathbb{N}$, $Y\in 
\ok{L_\boplus^i (\rho^\forall_M)}$ $\Lra$ $\ok{\bominus^i Y} = \ok{\rho^\forall_M
(\bominus^i Y)}$. Therefore, the thesis follows. \qedhere
\end{proof}

The following result provides a further useful characterization of the
complete core based on the structure
of the transition system.   We use the following notation:
given a transition system $\tuple{\mathbb{S},\sra}$ and states
$r,s\in \mathbb{S}$, for any $k>0$, $r\rak s$ iff $r=r_0 \sra
r_1 \shortrightarrow r_2 \shortrightarrow \ldots \shortrightarrow r_k =s$, where
$\{r_1,...,r_{k-1}\}\subseteq \mathbb{S}$. Moreover, we consider the
following property $\pr$ for any
$S\subseteq \mathbb{S}$:  
$$\pr(S) \text{~~~~iff~~~~} \exists k > 0, q\in S, r\in
\mathbb{S}\smallsetminus S, t\in\mathbb{S}.\; q \rak t \text{~and~}
r \rak t.$$

\begin{theorem}\label{car}
Let $M=\ma$, for some total transition system
$\tuple{\mathbb{S},\sra}$. Then, for any $S\subseteq
\mathbb{S}$, $\gamma_M^\forall (S) \not\in
\core_\boplus (\rho^\forall_{M})$ iff $\pr(S)$. 
\end{theorem}
\begin{proof}
($\Leftarrow$) 
Assume that there exist $k > 0$,  $q\in S$, $r\in
\mathbb{S}\smallsetminus S$, $t\in\mathbb{S}$ such that $q \rak t$ and
$r \rak t$. 
By Theorem~\ref{teosu}, it is enough to show that $\ok{\bominus^k
(\cup_{s\in S} M_{\downarrow s})} \supsetneq
\ok{\rho^\forall_{M} (\bominus^k ( \cup_{s\in S} M_{\downarrow s}))}$.
Since $q \rak t$ and 
$\tuple{\mathbb{S},\shortrightarrow}$ is total, there exists
$\tuple{j,\pi}\in M$ such that $\pi_j = q$ and
$\pi_{j+k}=t$. Since $q\in S$, we have that $\tuple{j,\pi}\in \cup_{s\in S}
M_{\da s}$ and therefore 
$\tuple{j+k,\pi}\in \ok{\bominus^k (\cup_{s\in S}
M_{\da s})}$. On the other hand,
since $r \rak t$ and $\tuple{\mathbb{S},\shortrightarrow}$ is total,
there exists $\tuple{l,\tau}\in M$ such 
that $\tau_l = r$ and $\tau_{l+k}= t = \pi_{j+k}$. Thus,
$\tuple{l+k,\tau} \in M_{\da \pi_{j+k}}$, while $\tuple{l+k,\tau} \not
\in \ok{\bominus^k (\cup_{s\in S} M_{\da s})}$ because $\tau_l = r \not \in
S$. Thus, by definition of $\rho_M^\forall$, 
this means that $\tuple{j+k,\pi} \not \in \rho^\forall_M
(\bominus^k (\cup_{s\in S} M_{\da s}))$. 
\\
($\Rightarrow$) By Theorem~\ref{teosu}, there exist $k>0$ and
$\tuple{j,\beta}$ such that (i)~$\tuple{j,\beta} \in \bominus^k
(\cup_{s\in S} M_{\da s})$ and
(ii)~$\tuple{j,\beta}\not \in \rho^\forall_{M} (\bominus^k 
(\bigcup_{s\in S} M_{\downarrow s}))$. Thus,
by (i),
$\tuple{j-k,\beta} \in \cup_{s\in S}
M_{\da s}$, i.e., $\beta_{j-k} \in S$. Moreover, by (ii), $M_{\da
\beta_j} \not \subseteq \bominus^k 
(\bigcup_{s\in S} M_{\downarrow s})$, so that there exists
$\tuple{l,\pi} \in M$ such that $\pi_l = \beta_j$ and $\tuple{l-k,\pi}
\not \in \cup_{s\in S} M_{\da s}$, i.e., $\pi_{l-k} \not \in S$. 
Summing up, we have that $\pi_{l-k} \rak
\pi_l$, $\beta_{j-k}\rak \pi_l$, $\pi_{l-k}\not \in S$ and
$\beta_{j-k}\in S$, that is $\pr(S)$. \qedhere
\end{proof}

Thus, by the characterization $(\ddagger)$ in Section~\ref{uca} of $\rho_M^\forall$ 
stating that $\{\gamma_M^\forall (S)\}_{S\subseteq \mathbb{S}}$ is the set of fixpoints
of $\rho_M^\forall$, the above result characterizes exactly
the fixpoints which must be
removed from 
$\rho^\forall_{M}$ in order to get the complete core
$\core_\boplus (\rho^\forall_{M})$.  
As an immediate consequence of Theorem~\ref{car}, observe 
that $M \in \core_\boplus (\rho^\forall_{M})$: in fact, by
Theorem~\ref{car}, $M = \gamma_M^\forall (\mathbb{S})$ and 
$\pr (\mathbb{S})$ is not satisfied. Let us also observe that $\pr
(S)$ holds iff $\pr(\bneg S)$ holds, so that $\gamma_M^\forall (S) \not
\in \core_\boplus (\rho^\forall_{M}) \: \Lra \: 
\gamma_M^\forall (\bneg S) \not
\in \core_\boplus (\rho^\forall_{M})$.

\begin{example}
Consider the transition system in Example~\ref{first}. We know that
$\rho_M^\forall = \{\gamma_M^\forall (\varnothing), \gamma_M^\forall
(\{1\}), \gamma_M^\forall (\{2\}), \gamma_M^\forall
(\{1,2\})\}$. Which elements are in   $\core_\boplus
(\rho^\forall_{M})$? $\gamma_M^\forall (\varnothing)$ and 
$\gamma_M^\forall (\{1,2\})$ always belong to $\ok{\core_\boplus
(\rho^\forall_{M})}$. Moreover, note that $1\rakn{1} 2$ and $2 \rakn{1}
2$ so that $\pr (\{1\})$ holds. Hence, by Theorem~\ref{car}, 
$\ok{\gamma_M^\forall (\{1\})}$ and $\ok{\gamma_M^\forall (\{2\})}$ do not
belong to $\ok{\core_\boplus
(\rho^\forall_{M})}$. 
\ddef
\end{example}

By exploiting the above constructive result, we are also able to
characterize the structure of transition systems whose models induce
a universal closure which is complete for
next-time. These are the transition systems $\tuple{\mathbb{S},\sra}$
such that $\sra$ is injective: the relation
$\rightarrow$ is \textit{injective} when $$\forall r,s,t\in
\mathbb{S}.\, (r\shortrightarrow t \;\:\&\;\: s \shortrightarrow t ) \Rightarrow
r=s.$$

\begin{theorem}\label{inj}
Let $M=\ma$, for some total transition system
$\tuple{\mathbb{S},\sra}$.
Then, $\rho^\forall_{M}$ is complete for $\boplus$ if and only if $\sra$ is
injective. 
\end{theorem}
\begin{proof}
$\rho^\forall_{M}$ is complete for $\boplus$ iff 
$\core_\boplus (\rho^\forall_{M}) = \rho^\forall_{M}$ iff 
$\core_\boplus (\rho^\forall_{M}) \sqsubseteq \rho^\forall_{M}$ iff 
$\rho^\forall_{M} \subseteq \core_\boplus (\rho^\forall_{M})$. Thus:\\
($\Rightarrow$) By hypothesis, for any $s\in \mathbb{S}$,  
$\gamma_M^\forall (\{s\}) \in \core_\boplus (\rho^\forall_{M})$. Thus, 
by Theorem~\ref{car}, for any $r,s,t\in \mathbb{S}$ such that $r\neq
s$, we have that  for any $k > 0$, $s\rak t$
implies $\neg (r \rak t )$.  
Hence, for any $r,s,t\in \mathbb{S}$ and for any $k>0$, 
$r \rak t$ and $s\rak t$ imply $s=r$. Therefore, for $k=1$, this
implies that $\sra$ is injective. \\
($\Leftarrow$) Let $\sra$ be injective. Let $r,s,t\in \mathbb{S}$ and
$k>0$ such that $r \rak t$ and $s\rak t$, i.e., $r \ra r_1 \sra \ldots
\sra r_{k-1}\sra t$ and $s \sra s_1 \sra \ldots \sra s_{k-1}\sra t$. Then,
by injectivity, $r_{k-1}=s_{k-1}$, and in turn, still by injectivity, 
$r_{k-2}=s_{k-2}$, and so on, so that we get $r=s$. Hence, 
for any $r,s,t\in \mathbb{S}$,
for any $k > 0$, $s\rak t$ and 
$r \rak t$ imply $r=s$. This means that, for any $s\in \mathbb{S}$,
$\ok{\pr (\{s\})}$ 
does not
hold. Thus, by 
Theorem~\ref{car}, $\ok{\gamma_M^\forall (\{s\})} \in \ok{\core_\boplus
(\rho^\forall_{M})}$. Since $\ok{\core_\boplus
(\rho^\forall_{M})}$ is a uco on $\ok{\wp(\mathbb{T})_\supseteq}$, its set of
fixpoints is closed under arbitrary set-unions. 
Moreover, since $\ok{\gamma_M^\forall}$ is co-additive on
$\ok{\wp(\mathbb{S})_\supseteq}$, we have that $\ok{\gamma_M^\forall}$ preserves
arbitrary set-unions. Thus, for any $S\subseteq \mathbb{S}$, 
$\ok{\gamma_M^\forall
(S)}=\ok{\cup_{s\in S}} \ok{\gamma_M^\forall(\{s\})} \in \ok{\core_\boplus
(\rho^\forall_{M})}$. Thus, since $\rfm = \{\ok{\gamma_M^\forall}
(S)\}_{S\subseteq \mathbb{S}}$, it turns out that $\ok{\rho_M^\forall}
\subseteq   \ok{\core_\boplus
(\rho^\forall_{M})}$. \qedhere
\end{proof}

It is worth noting that injectivity means that each computation step
is reversible, i.e.\ the reversed transition system
$\tuple{\mathbb{S},\shortleftarrow}$ obtained by reversing the transition
relation is deterministic.  This is the case of Bennett's reversible
computations \cite{Benn81}, i.e.\ computations whose output uniquely defines the
input, which have been extensively studied by many authors in different contexts.  
Let us also observe that if $s\in \mathbb{S}$ is
a stalling state, i.e.\ such that $s\sra s$, then the injectivity of
the transition relation requires that $t\not\ra s$ for any $t\neq s$,
i.e., $s$ cannot be reached by any other state so that $s$ must
necessarily be an
initial system state.

\begin{figure}
\begin{center}
\begin{tabular}{cc}
\mbox{
\xymatrix@=10pt{
 *+<25pt>[o][F]{red}\ar@{->}[rr] & & *+<20.5pt>[o][F]{green}\ar@{->}[rr]
       & & *+<17pt>[o][F]{yellow}\ar@(d,d){->}[llll] \\
}}~~~~~
&~~~~~
\mbox{
\xymatrix@=10pt{
 *+<15pt>[o][F]{red}\ar@{->}[rr]
       & & *+<17pt>[o][F]{go}\ar@(d,d){->}[ll] \ar@(ul,ur)[] \\
}}
\end{tabular}
\end{center}
\caption{A traffic light controller and its abstract version.}\label{fig}
\end{figure}
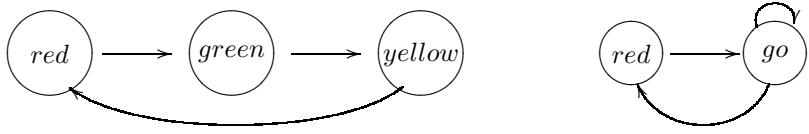

\begin{example}\label{pippa}
Consider a traffic light controller modelled by the transition system 
$\tuple{\mathbb{S},\sra}$
depicted in Figure~\ref{fig} generating the model $M$.   
Then,
$\tuple{\mathbb{S},\ra}$ is total and injective, and therefore, by
Theorem~\ref{inj}, the corresponding universal closure is
complete for next-time, so that
$\core_\boplus (\rho^\forall_{M}) = \rho^\forall_{M}$. \\  
Consider instead the abstract transition system
$\ok{\tuple{\mathbb{S}^{\sharp}=\{\mathit{red},\mathit{go}\},\sra^{\sharp}}}$ 
induced by the state
partition $\{\{ \mathit{red}\}, \{\mathit{green},\mathit{yellow}\}\}$
(see \cite{CGP99} for an introduction to abstract model checking)
and still depicted in Figure~\ref{fig}. 
In this case, $\ok{\tuple{\mathbb{S}^{\sharp},\sra^{\sharp}}}$ is total
but it is not injective. Let $\ok{M^\sharp}$ be the model generated by
$\ok{\tuple{\mathbb{S}^{\sharp},\sra^{\sharp}}}$. We exploit
Theorem~\ref{car} in order to compute
the complete core in this case. 
It turns out that $\ok{\mathit{red}\rat^\sharp
\mathit{go}}$ 
and $\ok{\mathit{go}\rat^\sharp \mathit{go}}$, so that 
$\ok{{\ensuremath{P_{{\scriptscriptstyle
\!\!\rightarrow^\sharp}}\!}}(\mathit{red})}$ and 
$\ok{{\ensuremath{P_{{\scriptscriptstyle
\!\!\rightarrow^\sharp}}\!}}(\mathit{go})}$ do not hold. 
Thus, in this case it turns out that 
the complete core is trivial, i.e., 
$\ok{\core_\boplus (\rho^\forall_{M^\sharp})} = \{\varnothing,
\ok{M^\sharp}\}$. \\
Let us also observe that any abstraction with at least two states of
$\tuple{\mathbb{S},\sra}$ induces an abstract transition system for
which the universal closure is not complete for next-time.
This is not always the case for abstract transition systems.  For
example, in the
case of an infinite counter modelled by a concrete transition system
$\tuple{\mathbb{S},\sra}$ where $\mathbb{S}=\mathbb{Z}$ and $x\sra y$
iff $y=x+1$, it turns out that both $\tuple{\mathbb{S},\sra}$ and the
abstract transition system $\tuple{\{\mathit{even},\, \mathit{odd}\},\sra^p}$ with
$\sra^p\ud \{\mathit{odd}\ra \mathit{even}, \mathit{even}\ra
\mathit{odd}\}$, obtained by the even/odd partition of integer numbers,  
are such that the corresponding universal closures are complete
for $\oplus$: in fact, both transition relations are injective
and therefore Theorem~\ref{inj} applies.  \ddef
\end{example}

\subsubsection{Complete shell}

By applying again Theorem~\ref{ft}, 
let us now characterize the set of fixpoints of the 
complete shell of the universal
closure for next-time.

\begin{theorem}\label{teocu}
The set of fixpoints of 
$\ok{\shell_\boplus (\rho^\forall_{M})}$ 
is $\Cl^\cup (\{\bominus^n (X)~|~ n\in \mathbb{N}, X\in
\rho_M^\forall \})$.
\end{theorem}
\begin{proof}
By Theorem~\ref{ft} and Remark~\ref{segu}, 
$\shell_\boplus (\rho^\forall_{M}) = \sqcap_{i\in \mathbb{N}} 
R_\boplus^i (\eta))$, where 
$R_\boplus (\eta) = 
\Cl^\cup (\{\cap \{X\in \wp(\mathbb{T})~|~\boplus X \supseteq Y\} ~|~ Y\in
\eta \}) = \Cl^\cup (\{ \bominus (Y) ~|~ Y\in \eta\})$. Moreover, 
for
any $i\in \mathbb{N}$,  $R_\boplus^i (\eta) = \Cl^\cup (\{ \bominus^i
(Y)~|~ Y\in \eta\})$.   
Thus, it turns out that 
\begin{align*}
\shell_\boplus (\rho^\forall_{M}) & = \sqcap_{i\in \mathbb{N}} R_\boplus^i
(\rho_M^\forall )\\
 & = \Cl^\cup (\cup_{i\in \mathbb{N}} \Cl^\cup (\{
\bominus^i (Y)~|~Y\in \rho^\forall_M\})) \\
& = \Cl^\cup (\cup_{i\in \mathbb{N}} \{
\bominus^i (Y)~|~Y\in \rho^\forall_M\}) \\
&= \Cl^\cup (\{\bominus^i (Y)~|~ i\in \mathbb{N},\, Y \in
\rho^\forall_M\}). \qedhere 
\end{align*}
\end{proof}

Thus, in order to
minimally refine the universal closure $\ok{\rho_M^\forall}$ to a complete
closure for the next-time $\boplus$, one must close the image of 
$\ok{\rho_M^\forall}$ under the application of the inverse of
$\boplus$, i.e., the previous-time trace operator $\bominus$. 

As a consequence of Theorem~\ref{teocu}, we can 
also provide a characterization  of
$\ok{\shell_\boplus (\rho^\forall_{M})}$ as a function. 
Given $\tuple{i,\sigma}\in \mathbb{T}$, $M\in \wp(\mathbb{T})$ and $k\in \mathbb{Z}$, 
let us define:
$$\ok{M^{k}_{\downarrow \tuple{i,\sigma}}} \ok{\ud} \ok{\{\tuple{j,\tau}\in
M~|~\tau_{j+k}=\sigma_{i+k}\}}.$$ 
This is a generalization of the
(current) state projection, since $M_{\downarrow \sigma_i} =
\ok{M^{0}_{\downarrow \tuple{i,\sigma}}}$. 
In particular, if $k\in \mathbb{N}$, $\ok{M^{-k}_{\downarrow
\tuple{i,\sigma}}}$ can be thought of as 
the $k$-th past state
projection of $M$. 

\begin{theorem}\label{main2}
$\ok{\shell_\boplus (\rho^\forall_{M})}=
 \lambda X. \{\tuple{i,\sigma}\in M~|~\exists k\in \mathbb{N}.\;
\ok{M^{-k}_{\downarrow \tuple{i,\sigma}}} \subseteq  X\}$.   
\end{theorem}
\begin{proof}
By Theorem~\ref{teocu}, we have that $\ok{\shell_\boplus
(\rho^\forall_{M})}
= \lambda X. \cup \{ \bominus^n (Z)~|~ n\in \mathbb{N}, \, Z\in \rfm,
\bominus^n (Z) \subseteq X\}$. Thus, let us show that for any
$X\subseteq \mathbb{T}$, 
$$\cup \{ \bominus^n (Z)~|~ n\in \mathbb{N}, \, Z\in \rfm,\,
\bominus^n (Z) \subseteq X\} = \{\tuple{i,\sigma}\in M~|~\exists k\in \mathbb{N}.\;
\ok{M^{-k}_{\downarrow \tuple{i,\sigma}}} \subseteq  X\}.$$
$(\subseteq)$: Let $\tuple{i,\sigma} \in \bominus^n (Z)$, for some
$n\in \mathbb{N}$ and  $Z\in \rfm$ such that
$\bominus^n (Z) \subseteq X$. Then, $\tuple{i-n,\sigma}\in Z$ and,
since $Z\in \rfm$, $\tuple{i-n,\sigma} \in M$. Let us show that 
$\ok{M^{-n}_{\downarrow \tuple{i,\sigma}}} \subseteq  X$. Consider
$\tuple{j,\tau}\in M$ such that $\tau_{j-n}=\sigma_{i-n}$. Since 
$\tuple{i-n,\sigma}\in Z$ and
$Z\in \rfm$, we have that $\tuple{j-n,\tau}\in Z$, so that $\tuple{j,\tau} \in
\bominus^n (Z)$. Hence,  $\bominus^n (Z) \subseteq X$ implies
$\tuple{j,\tau} \in X$. \\
$(\supseteq)$:
Consider $\tuple{i,\sigma} \in M$ such that $\ok{M^{-k}_{\downarrow
\tuple{i,\sigma}}} \subseteq  X$ for some $k\geq 0$. We consider
$M_{\downarrow \sigma_{i-k}}\in \rfm$ and we observe that
$\tuple{i,\sigma} \in \ok{\bominus^k (M_{\downarrow \sigma_{i-k}})}$. In
order to conclude, let
us check that $\ok{\bominus^k (M_{\downarrow \sigma_{i-k}})} \subseteq
X$. Consider $\tuple{j,\tau}\in \ok{\bominus^k (M_{\downarrow
\sigma_{i-k}})}$, so that $\tuple{j-k,\tau}\in \ok{M_{\downarrow
\sigma_{i-k}}}$. Hence, $\tau_{j-k}=\sigma_{i-k}$, so that
$\tuple{j,\tau} \in \ok{M^{-k}_{\downarrow \tuple{i,\sigma}}}\subseteq
X$, and therefore $\tuple{j,\tau}\in X$.
\end{proof}

Thus, for any $X\in \wp(\mathbb{T})$, $\ok{\shell_\boplus (\rho^\forall_{M})} (X)$ 
throws away from $X$ all those traces either which are not in $M$ or which 
are in $M$ but any past or current state of the trace does not universally satisfy
$X$.  
The intuition is that while the universal closure 
$\ok{\rho_M^\forall}$ considers present states
only (i.e., $\ok{M_{\downarrow \sigma_i}} \subseteq
X$), as expected, completeness for next-time forces to take into account any
past state (i.e., $\exists k\in \mathbb{N}.\;
\ok{M^{-k}_{\downarrow \tuple{i,\sigma}}} \subseteq  X$). Therefore, in
order to design a suitable abstract domain for representing
$\ok{\shell_\boplus (\rho^\forall_{M})}$ we need 
``to prolong the abstract domain $\wp (\mathbb{S})_\supseteq$ in the past'' as
follows. 

\begin{definition}
Define $\ok{\sv} \ud \,\ok{\zm \ra
\wp (\mathbb{S})}$, where $\ok{\zm}$ is the set of nonpositive integers. 
Observe that $\ok{\sv}$ is a complete lattice w.r.t.\ the standard pointwise ordering
$\ok{\dot{\supseteq}}$. \\
Given $z\in \zm$, $s\in\mathbb{S}$ and $M\in \wp(\mathbb{T})$, define
$\ok{M^z_{\downarrow s}\ud \,\{\tuple{i,\sigma}\in M~|~
\sigma_{i+z}=s\}}$. \\
The mappings $\ok{\alpha_{\forall_M}^\oplus}: \wp(\mathbb{T})\ra \sv$ and
$\ok{\gamma_{\forall_M}^\oplus}: \sv \ra \wp(\mathbb{T})$ are defined as follows:\\[5pt]   
\mbox{~~~~~}$\ok{\alpha_{\forall_M}^\oplus} (X)\ud \lambda z\in \zm. \: \{s\in
\mathbb{S}~|~ M^z_{\downarrow s} \subseteq X\};$\\[5pt]
\mbox{~~~~~}$\ok{\gamma_{\forall_M}^\oplus} (\Sigma)\ud \{\tuple{i,\sigma}\in M~|~
\exists k\in \nat.\: \sigma_{i-k}\in \Sigma(-k)\}$. \ddef
\end{definition}

\begin{corollary}\label{csm}
$\ok{(\alpha_{\forall_M}^\oplus, \wp(\mathbb{T})_\supseteq,
\sv_{\!\!\!\!\!\! \mbox{}^{\scriptscriptstyle \dot{\supseteq}}}, 
\gamma_{\forall_M}^\oplus)}$ is a GC, and additionally a GI when
$M=\ma$, for some total transition system
$\tuple{\mathbb{S},\sra}$, which induces 
the closure $\ok{\shell_\oplus (\rho^\forall_{M})}$.
\end{corollary}
\begin{proof}
The fact that 
$\ok{(\alpha_{\forall_M}^\oplus, \wp(\mathbb{T})_\supseteq,
\sv_{\!\!\!\!\!\! \mbox{}^{\scriptscriptstyle \dot{\supseteq}}}, 
\gamma_{\forall_M}^\oplus)}$ is a GC/GI follows easily from the GC/GI 
$\ok{(\alpha^\forall_{M}, \wp(\mathbb{T})_\supseteq,
\wp (\mathbb{S})_\supseteq, \gamma^\forall_{M})}$.
Moreover, 
observe that $\ok{\gamma_{\forall_M}^\oplus \circ
\alpha_{\forall_M}^\oplus}$ coincides with the characterization of
$\ok{\shell_\oplus (\rho^\forall_{M})}$
given by Theorem~\ref{main2}. 
\end{proof}

Hence, the state abstract domain $\ok{\wp (\mathbb{S})_\supseteq}$
needs to be refined to a domain of infinite sequences of sets of
states, namely the ``prolongation'' of $\ok{\gamma_M^\forall}$ in the
past.  We index the sequences $\Sigma\in \sv$ over $\zm$, so that for
any and $i\in \nat$, $\ok{\Sigma(-i)\in \wp (\mathbb{S})}$ is
reminiscent of a set of states at time $-i\in
\zm$. 

As a consequence, it is easy to design an abstract domain for representing
the complete shell of the universal closure for both next- and
previous-time. In fact, the prolongation of $\ok{\wp
(\mathbb{S})_\supseteq}$ both in the past and in the future leads to 
the GI
$\ok{(\alpha_{\forall_M}^{\scriptscriptstyle \pm}, \wp(\mathbb{T})_\supseteq,
\svo_{\!\!\!\!\!\! \mbox{}^{\scriptscriptstyle \dot{\supseteq}}}, 
\gamma_{\forall_M}^{\scriptscriptstyle \pm})}$, where: 
\\[5pt]   
\mbox{~~~~~}$\ok{\alpha_{\forall_M}^{\scriptscriptstyle \pm}} (X)\ud \lambda z\in \mathbb{Z}. \: \{s\in
\mathbb{S}~|~ M^z_{\downarrow s} \subseteq X\};$\\[5pt]
\mbox{~~~~~}$\ok{\gamma_{\forall_M}^{\scriptscriptstyle \pm}} (\Sigma)\ud \{\tuple{i,\sigma}\in M~|~
\exists k\in \mathbb{Z}.\: \sigma_{i+k}\in \Sigma(k)\}$.

\begin{example}
Let us consider again the two states transition system in
Example~\ref{first} and the formula $\oplus \!\ominus\! p \in \clm$, where
$p=\boldsymbol{\sigma}_{1}$. Observe that $\grass{\oplus \!\ominus\! p} =
\grass{p} = M_{\downarrow 1}$. The formula $\oplus \!\ominus\! p$ is not
branchable, namely the abstract semantics of $\oplus \!\ominus\! p$ induced by
$\rfm$ is not complete. In fact, $\ok{\alpha_M^\forall (\grass{\oplus
\!\ominus\! p})} = \{1\}$ while
$\ok{\grass{\oplus \!\ominus\!
p}_M^\forall} = \ok{\pret_\sra (\postt_\sra (\alpha_M^\forall
(M_{\downarrow 1})))} = \ok{\pret_\sra (\postt_\sra (\{1\}))} = 
\ok{\pret_\sra (\{1\})} = \varnothing$.  \\
Let us check that for the above abstract domain $\svo$ completeness
does hold.  
In this case, the abstract semantics is as follows: 
$\grass{\oplus \ominus p}_M^{\scriptscriptstyle \pm} = 
\ok{\alpha_{\forall_M}^{\scriptscriptstyle \pm}} \!\circ 
\boplus \circ \ok{\gamma_{\forall_M}^{\scriptscriptstyle \pm}} \!\circ 
\ok{\alpha_{\forall_M}^{\scriptscriptstyle \pm}} \!\circ 
\bominus \circ \ok{\gamma_{\forall_M}^{\scriptscriptstyle \pm}} \!\circ 
\ok{\alpha_{\forall_M}^{\scriptscriptstyle \pm}} (M_{\downarrow 1})$. 
Hence, we have the following equalities:
\[
\begin{array}{l}
\ok{\alpha_{\forall_M}^{\scriptscriptstyle \pm}} (M_{\downarrow 1})
(z) = \left\{ \begin{array}{cl}
\!\!\varnothing & \text{if $z < 0$}\\ 
\!\!\{1\} & \text{if $z \geq 0$} 
\end{array} \right.
\\[15pt]
\ok{\gamma_{\forall_M}^{\scriptscriptstyle \pm}} 
(\ok{\alpha_{\forall_M}^{\scriptscriptstyle \pm}} (M_{\downarrow 1}))  
= M_{\downarrow 1}
\\[10pt]
\bominus (\ok{\gamma_{\forall_M}^{\scriptscriptstyle \pm}} 
(\ok{\alpha_{\forall_M}^{\scriptscriptstyle \pm}} (M_{\downarrow 1}))) 
= M_{\downarrow 1} \cup \{\tuple{i,\sigma}\in M~|~ \sigma_i=2,\,
\sigma_{i-1} =1\}
\\[10pt]
\ok{\alpha_{\forall_M}^{\scriptscriptstyle \pm}} (\bominus (\ok{\gamma_{\forall_M}^{\scriptscriptstyle \pm}} 
(\ok{\alpha_{\forall_M}^{\scriptscriptstyle \pm}} (M_{\downarrow
1})))) (z)  
= \left\{ \begin{array}{cl}
\!\!\varnothing & \text{if $z < -1$}\\ 
\!\!\{1\} & \text{if $z \geq -1$} 
\end{array} \right.
\\[15pt]
\ok{\gamma_{\forall_M}^{\scriptscriptstyle \pm}} 
(\ok{\alpha_{\forall_M}^{\scriptscriptstyle \pm}} (\bominus (\ok{\gamma_{\forall_M}^{\scriptscriptstyle \pm}} 
(\ok{\alpha_{\forall_M}^{\scriptscriptstyle \pm}} (M_{\downarrow
1}))))) = M_{\downarrow 1} \cup \{\tuple{i,\sigma}\in M~|~ \sigma_i=2,\,
\sigma_{i-1} =1\}
\\[10pt]
\boplus (\ok{\gamma_{\forall_M}^{\scriptscriptstyle \pm}} 
(\ok{\alpha_{\forall_M}^{\scriptscriptstyle \pm}} (\bominus (\ok{\gamma_{\forall_M}^{\scriptscriptstyle \pm}} 
(\ok{\alpha_{\forall_M}^{\scriptscriptstyle \pm}} (M_{\downarrow
1})))))) = M_{\downarrow 1} 
\end{array}
\]
As a consequence, it turns out that 
$$\ok{\alpha_{\forall_M}^{\scriptscriptstyle \pm}} (\grass{\oplus\!
\ominus\! p}) = \ok{\alpha_{\forall_M}^{\scriptscriptstyle \pm}}
(M_{\downarrow 1} ) = \ok{\alpha_{\forall_M}^{\scriptscriptstyle \pm}}
(\boplus (\ok{\gamma_{\forall_M}^{\scriptscriptstyle \pm}} 
(\ok{\alpha_{\forall_M}^{\scriptscriptstyle \pm}} (\bominus (\ok{\gamma_{\forall_M}^{\scriptscriptstyle \pm}} 
(\ok{\alpha_{\forall_M}^{\scriptscriptstyle \pm}} (M_{\downarrow
1}))))))) = \grass{\oplus\! \ominus\! p}_M^{\scriptscriptstyle \pm}$$
namely completeness holds for this abstract domain.
\ddef
\end{example}

\subsection{Time reversal}

Let us now analyze the time reversal operator. The universal
abstraction for the reversed model $\cl M$ is
characterized as follows. Of course, notice that if $M$ is generated
by a transition system $\tuple{\mathbb{S},\sra}$ then $\cl M$ is the
model generated by the reversed transition system
$\tuple{\mathbb{S},\shortleftarrow}$. 

\begin{lemma}\label{serve}
$\rho^\forall_{^{{\cal} \!} M} =
\cl  \circ \rho^\forall_{M}\circ \cl$.
\end{lemma}
\begin{proof}
Let us show that $\ok{\cl
(\rho^\forall_{M} (\mbox{}^\curvearrowleft X))= \rho^\forall_{^{\cal
\!} M} (X)}$. Let
$\ok{\tuple{i,\sigma} \in \cl (\rho^\forall_{M}
(\mbox{}^\curvearrowleft X))}$. Then, $\ok{\cl \tuple{i,\sigma} \in
\rho^\forall_{M} (\mbox{}^\curvearrowleft X)}$, and therefore 
$\cl \tuple{i,\sigma} \in M$ and
$M_{\downarrow \sigma_i}\subseteq  \cl X$. This implies $\tuple{i,\sigma}
\in \cl M$ and $\cl (M_{\downarrow \sigma_i}) \subseteq  X$. 
Since $\cl (M_{\downarrow \sigma_i}) = (\cl M)_{\downarrow
\sigma_i}$, this means that $\ok{\tuple{i,\sigma} \in
\rho^\forall_{^{\cal \!} M} (X)}$.  On the other hand, the previous
implications actually are equivalences, and thus the reverse inclusion
simply follows by going backward.  \qedhere
\end{proof}

\subsubsection{Complete core} 

Theorem~\ref{ft} allows us here to show  that 
the complete core is given by those fixpoints of
$\rho^\forall_{M}$ which also belong to the
universal closure $\rho^\forall_{^{\cal \!} M}$ relative to the
reversed model $\cl M$. 

\begin{theorem}\label{teosu2}
The set of fixpoints of $\ok{\core_{\scriptscriptstyle \cal} (\rho^\forall_{M})}$  is
$\ok{\{Y \in \wp(\mathbb{T})~|~ Y ,\, \cl Y \in \rho^\forall_{M}
\}}$. Moreover, 
$\ok{\core_{\scriptscriptstyle \cal} (\rho^\forall_{M})} =
\ok{\rho^\forall_{M} \sqcup \rho^\forall_{^{\cal \!} M}}$.
\end{theorem}
\begin{proof}
By Theorem~\ref{ft} and Remark~\ref{segu}, we have that 
$\ok{\core_{\scriptscriptstyle\cal} (\rho^\forall_{M})} = \sqcup_{i\in \mathbb{N}}
\ok{L_{\scriptscriptstyle \cal}^i} (\rfm)$, where  
$L_{\scriptscriptstyle \cal} (\eta) = 
\{ Y \in \wp(\mathbb{T})~|~
\cap \{X\in \wp(\mathbb{T})~|~\cl X \supseteq Y\} \in \eta \}$. 
Since $\cl X
\supseteq Y \Leftrightarrow X \supseteq \cl Y$, we have that
$L_{\scriptscriptstyle \cal} (\eta) = 
\{ Y \in \wp(\mathbb{T})~|~ \cl Y \in \eta \}$. Thus, for any $j>0$,
$\ok{L_{\scriptscriptstyle \cal}^{2j}} (\rfm)= \rfm$ and 
$\ok{L_{\scriptscriptstyle \cal}^{2j+1}} (\rfm)= 
L_{\scriptscriptstyle \cal} (\rfm)$. Hence, 
$\sqcup_{i\in \mathbb{N}}
\ok{L_{\scriptscriptstyle \cal}^i (\rho^\forall_M)} =
\ok{\rho^\forall_M} \sqcup L_{\scriptscriptstyle \cal}
(\rfm) = 
\ok{\{Y \in \wp(\mathbb{T})~|~ Y ,\, \cl Y \in \rfm
\}}$. 
Moreover, let us observe that $\ok{\cl Y \in \rho^\forall_M} \Leftrightarrow
\ok{\rho^\forall_M (\cl Y) =\cl Y} \Leftrightarrow \ok{\cl (\rho^\forall_M (\cl
Y)) = Y}$. Thus, by Lem\-ma~\ref{serve}, 
$\ok{\cl Y \in \rho^\forall_M \Leftrightarrow Y \in \rho^\forall_{^{\cal
\!} M}}$, and thus we also have that
$\ok{\core_{\scriptscriptstyle \cal} (\rho^\forall_{M})} =
\ok{\rho^\forall_{M} \sqcup \rho^\forall_{^{\cal \!} M}}$.
\qedhere
\end{proof}

This allows us to give a characterization of the
transition systems that induce universal closures which are complete
for time reversal.  It turns out that these are the symmetric 
transition systems: a relation $\sra$ is symmetric when
$\forall r,s \in \mathbb{S}.\, r\!\sra\! s \;\Rightarrow\; s\! \sra\!
r$. This means that in symmetric transition systems  
any computation step is reversibile. 

\begin{corollary}\label{corma}
Let $M=\ma$ for some total transition system
$\tuple{\mathbb{S},\sra}$. Then, $\rho^\forall_{M}$ is complete
for $\cl$ if and only if $\sra$ is symmetric.
\end{corollary}
\begin{proof}
Let us first observe that $\sra$ is symmetric iff $M = \cl M$. Let us show that
$\ok{\rho^\forall_{^{\cal \!} M}} \sqsubseteq \ok{\rho^\forall_M}   \; \Rightarrow \; M =
\cl M$: we have that $\cl M =\ok{\rho^\forall_{^{\cal \!} M} (\mathbb{T})} \supseteq
\ok{\rho^\forall_M (\mathbb{T} )}= M$, and in turn, by applying $\cl$,
$M \supseteq  \cl M$,
that is $\cl M =  M$. 
Thus, $\ok{\rho^\forall_{^{\cal \!} M}} 
\sqsubseteq \ok{\rho^\forall_M}   \; \Leftrightarrow \; M =
\cl M$.
Moreover, by
Theorem~\ref{teosu2}, $\ok{\rho^\forall_{M}}$ is complete for $\cl$ iff
$\ok{\core_{\scriptscriptstyle \cal} 
(\rho^\forall_{M})} = \ok{\rho^\forall_{M}}$ iff $\ok{\rho^\forall_{^{\cal
\!} M}} \sqsubseteq \ok{\rho^\forall_M}$. Hence, this closes the proof. \qedhere  
\end{proof}

Thus, in practice, the 
universal closure is rarely complete for time reversal, since symmetry
is not a realistic condition for most systems.

\begin{example}
Consider the abstract counter and the abstract traffic
light controller in Example~\ref{pippa}.
The transition relations of both systems are
symmetric, so that, by
Corollary~\ref{corma}, 
the universal closure is complete for time reversal. This is not the case
of the concrete three-state traffic light controller, since the
transition relation is not symmetric.  Observe that the model
generated by this transition system is as follows: 
$$M=\{ \tuple{i,\cdots
\mathit{red}~\mathit{green}~\mathit{yellow}~\mathit{red}~\mathit{green}~\mathit{yellow}
\cdots}~|~ i\in \mathbb{Z}\}.$$
Thus, for any $Y\subseteq M$, $Y ,\, \cl Y \in
\rho^\forall_{M}$ holds if and only if $Y=\varnothing$. Therefore, by 
Theorem~\ref{teosu2}, $\ok{\core_{\scriptscriptstyle\cal} (\rho^\forall_{M})} = \{
\varnothing\}$, i.e., the complete core is the trivial abstract domain
representing no information.
\ddef
\end{example}

\subsubsection{Complete shell}

Let us now apply our constructive approach to characterize the complete shell.

\begin{theorem}\label{asing}
The set of fixpoints of $\ok{\shell_{\scriptscriptstyle\cal} (\rho^\forall_{M})}$  is
$\ok{\Cl^\cup (\rho_M^\forall \cup 
\{Y \in \wp(\mathbb{T}) ~|~ \cl Y \in \rho^\forall_{M}
\})}$. Moreover, 
$\ok{\shell_{\scriptscriptstyle\cal} (\rho^\forall_{M})} =
\ok{\rho^\forall_{M} \sqcap \rho^\forall_{^{\cal \!} M}}$.
\end{theorem}
\begin{proof}
By Theorem~\ref{ft} and Remark~\ref{segu},
$\ok{\shell_{\scriptscriptstyle \cal} (\rho^\forall_{M})} = 
\sqcap_{i\in \mathbb{N}}
\ok{R_{\scriptscriptstyle\cal}^i (\rho^\forall_M)}$, where 
$R_{\scriptscriptstyle\cal} (\eta) = 
\Cl^\cup (\{\cap \{X\in \wp(\mathbb{T})~|~\cl X \supseteq Y\} ~|~ Y\in
\eta \}) = \Cl^\cup (\{ \cl Y ~|~ Y\in \eta\}) = \Cl^\cup (\{ Y ~|~
\cl Y \in \eta\})$. 
Since $\cl$ preserves arbitrary unions, for any $j>0$,
$\ok{R_{\scriptscriptstyle\cal}^{2j}} (\ok{\rho^\forall_M})= \ok{\rho^\forall_M}$ and 
$\ok{R_{\scriptscriptstyle\cal}^{2j+1}} (\ok{\rho^\forall_M})= 
\ok{R_{\scriptscriptstyle\cal} (\rho^\forall_M)}$. 
Hence, 
$\sqcap_{i\in \mathbb{N}}
\ok{R_{\scriptscriptstyle\cal}^i (\rho^\forall_M)} =
\ok{\rho^\forall_M \sqcap R_{\scriptscriptstyle\cal} 
(\rho^\forall_M)} = 
\ok{\Cl^\cup (\rho_M^\forall \cup 
\{ Y ~|~ \cl (Y) \in \rho^\forall_{M}
\})}$. Moreover, as observed in the proof of Theorem~\ref{teosu2}, 
$\ok{\cl Y \in \rho^\forall_M} \Leftrightarrow
\ok{\cl (\rho^\forall_M (\cl
Y)) = Y}$, and therefore, by Lem\-ma~\ref{serve}, 
$\ok{R_{\scriptscriptstyle\cal} 
(\rho^\forall_M)} = \ok{\rho^\forall_{^{\cal \!} M}}$, so that
we obtain that 
$\ok{\shell_{\scriptscriptstyle\cal} (\rho^\forall_{M})} =
\ok{\rho^\forall_{M} \sqcap \rho^\forall_{^{\cal \!} M}}$.
\end{proof}

It is therefore simple to design an abstract domain for representing
this complete shell. We consider the abstract domain $\wp
(\mathbb{S})^2_\supseteq$ as related to the concrete domain 
$\wp(\mathbb{T})_\supseteq$ by the
following abstraction and concretization maps:
\begin{itemize}
\item[] $\alpha_{\forall_M}^{\scriptscriptstyle \cal}\ud \lambda X.\tuple{\alpha_M^\forall
(X),\alpha_{^{\cal \!} M}^\forall(X) }$;
\item[] $\gamma_{\forall_M}^{\scriptscriptstyle\cal}\ud 
\lambda \tuple{X_1,X_2}. \gamma_M^\forall
(X_1)\cup \gamma_{^{\cal \!} M}^\forall (X_2)$. 
\end{itemize}
 
\noindent
As a consequence of Theorem~\ref{asing}, 
it turns out
$\ok{\shell_{\scriptscriptstyle \cal} (\rho^\forall_{M})}$ 
 is the closure induced by the  GI 
$(\ok{\alpha_{\forall_M}^{\scriptscriptstyle\cal}},
\ok{\wp(\mathbb{T})_\supseteq}, \ok{\wp
(\mathbb{S})^2_\supseteq},\ok{\gamma_{\forall_M}^{\scriptscriptstyle\cal}})$. 
Thus, the above result tells us that completeness for 
time reversal requires an additional component taking into account the
universal  abstraction for the reversed model $\cl M$.  

\subsection{Disjunction}\label{secdisj}
Finally, let us consider disjunction, namely set-union in the concrete
domain 
$\wp(\mathbb{T})$. 


\subsubsection{Complete core} 

\begin{theorem}\label{dsing}
$\ok{\core_\cup (\rho^\forall_{M})} = \lambda X.\varnothing$. 
\end{theorem}
\begin{proof}
By Theorem~\ref{ft} and Remark~\ref{segu}, we have that 
$\ok{\core_\cup (\rho^\forall_{M})} = \sqcup_{i\in \mathbb{N}}
L_{\cup}^i (\rho^\forall_M)$, where  
$L_\cup (\eta) = \{ Y \in \wp(\mathbb{T})~|~ \{ \cap \{ Z \in \wp(\mathbb{T})~|~
Z \cup X \supseteq Y\}\}_{X\in \wp(\mathbb{T})} \subseteq \eta \}$.  
Note
that, for any $X,Y\in \wp(\mathbb{T})$, $\cap \{ Z \in \wp(\mathbb{T})~|~ Z \cup
X \supseteq Y\} = Y \cap \bneg X$ and
$\downarrow\! Y \ud \{ Z\in \wp(\mathbb{T}) ~|~ Z \subseteq Y\} = \{Y \cap \bneg X
~|~ X\in\wp(\mathbb{T})\}$. 
Thus, $L_\cup (\eta) = \{ Y \in
\wp(\mathbb{T})~|~ \downarrow\! Y \subseteq \eta\}$.  Also, let us observe 
that $L_\cup (\eta) \subseteq \eta$ and 
$\downarrow \! L_\cup (\eta)= L_\cup (\eta)$, so that, for any $i\geq
2$, $L_{\cup}^i (\rho^\forall_M) = L_{\cup} (\rho^\forall_M)$, and therefore
$\sqcup_{i\in \mathbb{N}}
L_{\cup}^i (\rho^\forall_M) = \{ Y \in \wp(\mathbb{T})~|~ \downarrow\! Y \subseteq
\rho^\forall_M\}$. 
Consider now some $Y\in \wp(\mathbb{T})$ 
such that $\downarrow\! Y \subseteq
\rho^\forall_M$. 
Then, $Y\in \rho_M^\forall$,  so that there exists some $S\subseteq
\mathbb{S}$ such that $Y=\gamma_M^\forall (S)$. If $s\in S$ then there
exists some $\tuple{i,\sigma}\in M_{\da s} \subseteq \gamma_M^\forall
(S)$, so that $\{\tuple{i,\sigma}\} \subseteq Y$. It turns out that
$\{\tuple{i,\sigma}\} \not \in \rho_M^\forall$ because
$\gamma_M^\forall (\{\sigma_i\}) = M_{\da \sigma_i}$ and, by
Hypothesis~\ref{hyppo}~(i), $|M_{\da \sigma_i}|>1$. This means that 
if $S\neq \varnothing$ then $\downarrow \! Y \not \subseteq
\rho_M^\forall$. Thus, $\ok{\core_\cup (\rho^\forall_{M})} =
\{\varnothing\}$, i.e., the core
is the top closure $\lambda X.\varnothing$.  \qedhere
\end{proof}

The greatest closure $\lambda X.\varnothing$ represents the
straightforward uninformative abstract domain consisting of a unique
abstract value which is the abstraction of any concrete value.  The
above result states that there is no 
further abstraction, but for the straightforward abstraction, 
of the universal abstraction 
which is complete for disjunction. As a consequence, we will 
prove later that any abstraction, but for the
straightforward one, of the state-based model checking for a
temporal calculus that includes an unrestricted connective of disjunction
is incomplete for the trace-based semantics.

\subsubsection{Complete shell}

\begin{theorem}\label{dsing2}
$\ok{\shell_\cup (\rho^\forall_{M})} = \lambda X. X\cap M$, so that
the set of fixpoints of $\ok{\shell_\cup (\rho^\forall_{M})}$
 is $\{X \in
\wp(\mathbb{T})~|~ X\subseteq M\}$. 
\end{theorem}
\begin{proof}
By Theorem~\ref{ft} and Remark~\ref{segu}, 
$\ok{\shell_\cup (\rho^\forall_{M})} = \ok{\sqcap_{i\in \mathbb{N}}}
\ok{R_{\cup}^i (\rho^\forall_M)}$, where
$\ok{R_\cup (\eta)} = \ok{\Cl^\cup (\{\cap
\{X\in\wp(\mathbb{T})~|~ X\cup Y\supseteq Z\}\}_{Y\in \wp(\mathbb{T}),\, Z \in
\eta} )} = \ok{\Cl^\cup 
(\{Z\cap \bneg Y ~|~ Y\in \wp(\mathbb{T}),\: Z \in
\eta\})}  = \ok{\Cl^\cup 
(\{Z\cap Y ~|~ Y\in \wp(\mathbb{T}),\: Z \in
\eta\})}$. Thus, we have  that $\sqcap_{i\in \mathbb{N}}
\ok{R_{\cup}^i (\rho^\forall_M)} = \ok{R_\cup (\rho_M^\forall)}$. It remains to 
observe that $\ok{\Cl^\cup} 
\ok{(\{Z\cap Y ~|~ Y\in \wp(\mathbb{T}),\: Z \in
\eta\})} = \ok{\{X \in \wp(\mathbb{T})~|~
X \subseteq M \}}$: this is an immediate set-theoretic consequence of the
fact that $M\in \rfm$ and that if $Z\in
\rfm$ then $Z \subseteq M$. Moreover, let us also note that 
the set of
fixpoints of $\lambda X. X\cap M$ is $\{X \in \wp(\mathbb{T})~|~
X\subseteq M \}$. 
\end{proof}

As a consequence, let us also notice that 
$\ok{\shell_\cup (\rho^\forall_{M})}$ 
is the closure induced by the  GI 
$\ok{(\alpha_{\forall_M}^\cup, \wp(\mathbb{T})_\supseteq,\wp
(M)_\supseteq,\gamma_{\forall_M}^\cup)}$, where
$\ok{\alpha_{\forall_M}^\cup} \ud \lambda X.X\cap M$ and
$\ok{\gamma_{\forall_M}^\cup} \ud \lambda X. X $. 
Hence, 
the complete shell
of the universal abstraction for the union
is ``essentially'' the identity mapping.  More precisely, for a given
model $M$, the closure 
$\ok{\shell_\cup (\rho^\forall_{M})}$ can be represented by the
abstract domain $\wp (M)_\supseteq$ endowed with the
abstraction map $\lambda X.X\cap M$ which simply removes those traces
which are not in $M$. This means  that completeness
for disjunction indeed requires all the traces in $M$.

Once again the above complete shell was characterized by exploiting the
constructive method in Section~\ref{ccs}. This complete shell can be also
obtained in a noncostrutive way.\footnote{This has been suggested by
one anonymous referee.} 
\begin{lemma}\label{reflem}
Let $X$ be any set and $\rho \in \uco(\wp(X)_\supseteq)$ such that
$\rho(M)=M$.  If $\rho$ is finitely additive then for any $Z\subseteq M$, $\rho(Z)=Z$.
\end{lemma}
\begin{proof}
Assume by contradiction that $Z\subseteq M$ is such that
$\rho(Z)\subsetneq Z$, and let $x\in Z \smallsetminus \rho(Z)$. Then,
$x\not\in M\smallsetminus Z$, so that $x\not \in \rho(M\smallsetminus
Z)$. Moreover, since $\rho(M\cap Z) \subseteq \rho(Z)$, we also have
that $x\not \in \rho(M\cap Z)$.
On the other hand, $x\in M=\rho(M)=\rho((M\cap Z) \cup M\smallsetminus
Z)$, so that $\rho(M\cap Z)\cup \rho(M\smallsetminus Z) \subsetneq \rho((M\cap
Z) \cup (M\smallsetminus Z))$, i.e., $\rho$ is not additive, a contradiction.
\end{proof}

Let us observe that $\rho \in \uco(\wp(\mathbb{T})_\supseteq)$ is
complete for finite set-union when for any $X,Y\in \wp(\mathbb{T})$,
$\rho(X\cup Y) = \rho(\rho(X)\cup \rho(Y))= \rho(X)\cup \rho(Y)$, that
is, when $\rho$ is finitely additive. This observation allows us to
show that $\ok{\shell_\cup
(\rho^\forall_{M})} = \lambda X. X\cap M$ in a nonconstructive way:
by Lemma~\ref{reflem}, since $M\in \rfm \subseteq \ok{\shell_\cup
(\rho^\forall_{M})}$, it turns out that for any $X\subseteq M$, $X\in 
\ok{\shell_\cup (\rho^\forall_{M})}$; hence, 
$\{X \in \wp(\mathbb{T})~|~ 
X\subseteq M \} \subseteq \ok{\shell_\cup (\rho^\forall_{M})}$, and
since $\{X \in \wp(\mathbb{T})~|~ 
X\subseteq M \}$ is (the set of fixpoints of) the
closure $\lambda X. X\cap M$ which is finitely additive, i.e.\
complete for set-union, we have that $\ok{\shell_\cup
(\rho^\forall_{M})} = \lambda X. X\cap M$. 
Let us remark that in this easy nonconstructive proof one
first needs to guess some abstract domain and then to prove that this
is indeed the complete shell. By contrast, our proof is easy as well
and, more importantly, 
constructive so that it is enough to apply
the methodology in Section~\ref{ccs} to characterize the
complete shell.

\subsection{All the connectives}\label{atc}
To conclude our analysis, 
let us characterize the complete core and shell of the universal
checking closure for all the connectives of the $\mus$-calculus, i.e.,
the set $\mathrm{TT}$ of all the trace transformers. 
We need to take
care of the following technicality.  As far as the universal
quantifier is concerned, the following restriction is
needed.  We just consider the unary restrictions $\lambda X. \fab
(N,X):\wp(\mathbb{T})\ra\wp(\mathbb{T})$, where $N\subseteq M\cup \cl
M$, because the binary
trace transformer $\fab :\wp(\mathbb{T}) \times \wp(\mathbb{T}) \ra\wp(\mathbb{T})$ is
neither monotone nor antitone in its first argument, while 
given any $N\in \wp(\mathbb{T})$, the
unary restriction $\lambda X. \fab (N,X)$ is instead monotone. 
Standard universal 
quantification can be expressed, because, 
as recalled  in
Section~\ref{tai},  
$\ok{\boldsymbol{\forall} \phi \ud
\boldsymbol{\forall} \: (\gabba \boldsymbol{\pi}_\sra) : \phi}$,
where $\grass{\,\gabba (\boldsymbol{\pi}_\sra)} = \mtr$.
In the sequel, we will use the following compact
notation: $M^* \ud M\cup\, \cl M$.
Hence, the set of trace transformers of the $\mus$-calculus is
$\mathrm{TT}\ud \ok{\{\boldsymbol{\sigma}_S\}_{S\in \wp(\mathbb{S})}}$ $\cup$ 
$\ok{\{\boldsymbol{\pi}_t\}_{t\in \wp (\mathbb{S}^2)}}$ $\cup$ $\ok{\{
\boplus,\cup,\bneg,\cl \}\cup\{\lambda X.\fab(N,X)\}_{N\subseteq
M^*}}$. 
As $\mathrm{TT}$ includes negation which is antimonotone, observe that
the existence of the complete core and shell of the universal closure
for all the connectives is not guaranteed.  
However, since the complete core of $\rfm$ for negation and disjunction is the greatest
closure  $\lambda X.\varnothing$ (by
Theorems~\ref{coreneg} and ~\ref{dsing}), as a straight consequence 
we obtain that $\lambda X.\varnothing$ is also the
complete core of $\rho^\forall_{M}$ for the set $\mathrm{TT}$ of trace
transformers, that is $\ok{\core_{\mathrm{TT}} (\rho^\forall_{M})} = 
\lambda X. \varnothing$. 
On the other hand, the complete shell for all the connectives does
exist and is as
follows. 

\begin{theorem}\label{lconcbis}
$\ok{\shell_{\mathrm{TT}} (\rho^\forall_{M})} = 
\lambda X. X\cap M^*$, so that  the set of fixpoints of 
$\ok{\shell_{\mathrm{TT}} (\rho^\forall_{M})}$
is 
$\{X \in
\wp(\mathbb{T})~|~ X\subseteq M^*\}$. 
\end{theorem}
\begin{proof}
Let $\rho= \lambda X. X\cap M^*$ and note that this
is a closure on $\wp(\mathbb{T})_\supseteq$. 
The following points  show that $\rho \in
\Gamma(\wp(\mathbb{T})_\supseteq,\mathrm{TT})$.\\
(1) $\rho\in\Gamma(\wp(\mathbb{T})_\supseteq,
\{\boldsymbol{\sigma}_S\}_{S\in \wp(\mathbb{S})}\cup
\{\boldsymbol{\pi}_t\}_{t\in \wp (\mathbb{S}^2)} )$ because $\boldsymbol{\sigma}_S$ and
$\boldsymbol{\pi}_t$ are 0-ary operators.\\
(2) $\rho\in \Gamma(\wp(\mathbb{T})_\supseteq,\boplus)$. 
Since $\boplus$ preserves unions and intersections, 
given $X\in \wp(\mathbb{T})$, $\rho (\boplus
(\rho (X))) = \rho (\boplus (X) \cap (\boplus
( M) \cup \boplus(\cl (M)))) = \boplus (X) \cap (\boplus (M)\cup \boplus (\cl (M))) \cap
(M\cup \cl (M))$. Also, by
Hypothesis~\ref{hyppo}~(ii), 
$\boplus (M) = M$ and $\boplus (\cl (M))= \cl (M)$, and therefore
$\rho (\boplus
(\rho (X))) =  \boplus (X)   \cap
(M\cup \cl (M))= \rho (\boplus (X))$. \\
(3) $\rho \in
\Gamma(\wp(\mathbb{T})_\supseteq,\cup)$.  In fact,
$\rho (\rho (X)\cup \rho (Y)) =
\rho ((X\cap M^*) \cup (Y\cap M^*))= 
\rho ((X\cup Y)\cap M^*)=  
(X\cup Y)\cap M^* = \rho (X\cup Y)$. \\ 
(4) $\rho\in
\Gamma(\wp(\mathbb{T})_\supseteq,\bneg)$.
In fact,  $\rho (\bneg \rho (X)) = 
(\bneg (X\cap M^*))\cap M^* =  ((\bneg X)\cap
M^*) \cup ((\bneg M^*)\cap M^*) = 
(\bneg X)\cap
M^* = \rho (\bneg X)$. \\
(5) $\rho\in
\Gamma(\wp(\mathbb{T})_\supseteq,\cl)$. As $\cl$ preserves intersections
and, by Hypothesis~\ref{hyppo}~(ii), $\cl
(M^*)=M^*$, we have that $\rho (\cl (\rho
(X)))= \rho (\cl (X\cap M^*))= \rho (\cl (X)\cap
M^*)= \cl (X)\cap M^*= \rho (\cl (X))$.  
\\
(6) $\rho \in \Gamma(\wp(\mathbb{T})_\supseteq, \{\lambda
X.\fab(N,X)\}_{N\subseteq M})$. Let $N\subseteq M$ and $X\in
\wp(\mathbb{T})$, and  observe that for any $\tuple{i,\sigma}\in N$, we have
that $ N_{\downarrow \sigma_i}\subseteq X\cap (M^*) \Leftrightarrow
(N_{\downarrow \sigma_i}\subseteq X)$. Thus, $\rho  (\fab
(N,\rho(X)))= \{\tuple{i,\sigma}\in N~|~N_{\downarrow
\sigma_i}\subseteq X\cap M^*\} \cap M^* = \{\tuple{i,\sigma}\in
N~|~N_{\downarrow \sigma_i}\subseteq X\} \cap M^* =
\rho (\fab (N,X))$.\\
To conclude, consider any 
$\eta\in \uco(\wp(\mathbb{T})_\supseteq)$ such that 
$\eta \in
\Gamma(\wp(\mathbb{T})_\supseteq,\mathrm{TT})$ and $\eta \sqsubseteq
\rho^\forall_M$. 
Since 
$\eta \in
\Gamma(\wp(\mathbb{T})_\supseteq,\cup)$, by Theorem~\ref{dsing2}, we have that $\eta
\sqsubseteq \ok{\shell_\cup (\rho^\forall_{M})} = \lambda X. X \cap
M$. 
Moreover, $\eta \in
\Gamma(\wp(\mathbb{T})_\supseteq,\cl)$, and hence, by Theorem~\ref{asing},
$\eta \sqsubseteq \ok{\shell_\cal (\rho^\forall_{M})} =
\ok{\rho^\forall_{M}} \sqcap \ok{\rho^\forall_{^{\cal \!} M}} \sqsubseteq 
\ok{\rho^\forall_{^{\cal \!} M}}$. 
Thus, because $\eta \sqsubseteq 
\ok{\rho^\forall_{^{\cal \!} M}}$ and $\eta
\in \Gamma(\wp(\mathbb{T})_\supseteq,\cl)$, we have that $\eta \sqsubseteq 
\ok{\shell_\cup ( \rho^\forall_{^{\cal \!} M})}$. 
By Theorem~\ref{dsing2}, 
$\ok{\shell_\cup ( \rho^\forall_{^{\cal \!} M})} = \lambda X. X \cap
\cl (M)$, so that $\eta \sqsubseteq \lambda X. X \cap
\cl (M)$. Hence, we obtained that $\eta \sqsubseteq 
(\lambda X. X \cap M) \sqcap (\lambda X. X \cap
\cl (M))= \rho$. Thus, $\ok{\shell_{\mathrm{TT}} (\rho^\forall_{M})} =
\rho$. 
\end{proof}

Let us observe that $\wp (M^*)_\supseteq$ is a suitable
abstract domain for representing this complete shell because the  GI 
$(\alpha_{\forall_M}, \wp(\mathbb{T})_\supseteq,\wp
(M^*)_\supseteq,\gamma_{\forall_M})$, where
$\alpha_{\forall_M}\ud$  $\lambda X.X\cap M^*$ and
$\gamma_{\forall_M} \ud \lambda X. X $, induces the closure 
$\lambda X. X\cap M^*$. The abstract domain $\wp (M^*)$ therefore
represents the traces of the system $\tuple{\mathbb{S},\ra}$ and of
the  reversed system $\tuple{\mathbb{S},\leftarrow}$.  

Let us remark that by exploiting the above results in
Sections~\ref{secneg}-\ref{secdisj}, it is not hard to
characterize the complete shell of the universal abstraction for any
subset of trace transformers. For example, when we leave out the
reversal operator from $\mathrm{TT}$, as one expects, it is easy to
show that in this case $\ok{\shell_{\mathrm{TT}} (\rho^\forall_{M})}=\lambda X. X\cap M$.

\section{Completeness of temporal languages}

Let  $\mathit{Op}$ be any set of temporal connectives, where each $\mathit{op}
\in \mathit{Op}$ has a corresponding arity $\sharp (\mathit{op})\geq
0$ so that constants are viewed as connectives whose  arity is 0. 
Following Cousot and Cousot \cite[Section~8]{CC00}, $\mathit{Op}$
 induces a corresponding
fixpoint temporal language $\mathfrak{L}_{\mathit{Op}}$ which is
inductively defined as
follows:
$$\mathfrak{L}_{\mathit{Op}}\ni \phi ::= X ~|~
\mathit{op}(\phi_1,...,\phi_n)~|~   \boldsymbol{\mu} X.\phi 
~|~   \boldsymbol{\nu} X.\phi$$ 
where  
$X\in \mathbb{X}$ and $\mathit{op}
\in \mathit{Op}$. Given any  set of states  $\mathbb{S}$ which
determines a corresponding set of traces $\mathbb{T}$,
the semantics of any connective $\mathit{op}$ with
arity $n\geq 0$ is given by a corresponding trace transformer 
$\boldsymbol{\mathit{op}}: \wp(\mathbb{T})^n \ra \wp(\mathbb{T})$. The set
of trace transformers that provide the semantics of connectives in
$\mathit{Op}$ is denoted by $\boldsymbol{\mathit{Op}}$. Hence, this
determines a trace semantics of $\mathfrak{L}_{\mathit{Op}}$, namely 
$\ok{\grass{\cdot}} : \mathfrak{L}_{\mathit{Op}} \ra \mathbb{E} \ra
\wp(\mathbb{T})$, which is inductively (and, possibly, partially due
to fixpoints) defined  as follows:
$$
\begin{array}{ll}
\grass{X} \xi = \xi (X) &
~~\grass{\boldsymbol{\mu} X.\phi}\xi = \lfp (\lambda
N\in\wp(\mathbb{T}).\grass{\phi}\xi[X/N])
\\[5pt]
\grass{\mathit{op} (\phi_1,...,\phi_n)} \xi =
\boldsymbol{\mathit{op}} (\grass{\phi_1}\xi,...,\grass{\phi_n}\xi) &
~~\grass{\boldsymbol{\nu} X.\phi}  \xi = \gfp (\lambda
N\in\wp(\mathbb{T}). \grass{\phi}\xi[X/N])
\end{array} 
$$

Thus, any abstraction of the concrete domain $\wp(\mathbb{T})$ induces
an abstract semantics for $\mathfrak{L}_{\mathit{Op}}$.
As described in Section~\ref{sbas}, the universal 
abstraction provides an example: the state semantics $\ok{\grass{\cdot}_M^\forall}$ is 
the abstract semantics induced
by $\rfm\in \uco(\wp(\mathbb{T}_\supseteq))$. 
In general, any 
abstract domain $\rho\in \uco(\wp(\mathbb{T})_\supseteq)$
induces the set of abstract environments $\mathbb{E}^\rho \ud
\mathbb{X}\ra \rho$. Hence, the abstract semantics
$\ok{\grass{\cdot}^\rho}:\mathfrak{L}_{\mathit{Op}} \ra
\mathbb{E}^\rho\ra \rho$ 
is defined as follows: 
$$
\begin{array}{ll}
\!\!\grass{X}^\rho \chi = \chi (X) &
\!\!\!\!\grass{\boldsymbol{\mu} X.\phi}^\rho \chi = \lfp (\lambda
N\in\rho.\grass{\phi}^\rho \chi[X/N])
\\[5pt]
\!\!\grass{\mathit{op} (\phi_1,...,\phi_n)}^\rho \chi =
\rho(\boldsymbol{\mathit{op}} (\grass{\phi_1}^\rho
\chi,...,\grass{\phi_n}^\rho \chi)) &
\!\!\!\!\grass{\boldsymbol{\nu} X.\phi}^\rho  \chi = \gfp (\lambda
N\in\rho. \grass{\phi}^\rho\chi[X/N])
\end{array} 
$$
Given a concrete
environment $\xi \in \mathbb{E}$, $\ok{\dot{\rho}(\xi)\ud \lambda
X.\rho(\xi (X))\in \mathbb{E}^\rho}$ is the corresponding abstract
environment induced by $\rho$.  
Soundness of $\rho$ for the language
$\mathfrak{L}_{\mathit{Op}}$ 
means that the abstract semantics $\ok{\grass{\cdot}^\rho}$ is sound, namely 
for any $\phi\in \mathfrak{L}_{\mathit{Op}}$ and
 $\ok{\xi\in \mathbb{E}}$, $\ok{\rho (\grass{\phi}\xi)} \subseteq
\ok{\grass{\phi}^\rho \dot{\rho} (\xi)}$. 
Completeness of $\rho$ for $\mathfrak{L}_{\mathit{Op}}$ means 
that
equality always holds.
As usual, the
abstract interpretation approach always ensures soundness, while 
completeness in general does not hold.

Given $\rho\in
\uco(\wp(\mathbb{T})_\supseteq)$, the complete shell of $\rho$
for $\mathfrak{L}_{\mathit{Op}}$, when it exists, is the most
abstract domain $\Shell_{\mathfrak{L}_{\mathit{Op}}} (\rho) 
\in \uco(\wp(\mathbb{T})_\supseteq)$ such that
$\Shell_{\mathfrak{L}_{\mathit{Op}}} (\rho)  
\sqsubseteq \rho$ and $\Shell_{\mathfrak{L}_{\mathit{Op}}} (\rho) $ is complete for 
$\mathfrak{L}_{\mathit{Op}}$. Complete cores for
$\mathfrak{L}_{\mathit{Op}}$ are defined dually. 

We recalled in Section~\ref{ccs} that if $\rho$ is complete for some
function $f$ then $\rho$ is also fixpoint complete for $f$.
Thus, as a straight consequence we obtain that
if $\rho\in
\uco(\wp(\mathbb{T})_\supseteq)$ is complete for 
$\boldsymbol{\mathit{Op}}$ and either $\rho$ does not contain
infinite descending chains or $\rho$ is co-continuous then $\rho$ is complete for 
$\mathfrak{L}_{\mathit{Op}}$. 
Moreover, it turns out that
complete shells and cores for a temporal language
$\mathfrak{L}_{\mathit{Op}}$ 
coincide with
complete shells and cores for the corresponding set 
$\boldsymbol{\mathit{Op}}$
of trace transformers.

\begin{theorem}\label{teofix}
Let $\rho\in
\uco(\wp(\mathbb{T})_\supseteq)$. 
If $\:\Shell_{{\boldsymbol{\mathit{Op}}}} (\rho)$ exists and  either 
does not contain
infinite descending chains or is co-continuous then 
$\Shell_{\mathfrak{L}_{\mathit{Op}}} (\rho) =
\Shell_{{\boldsymbol{\mathit{Op}}}} (\rho)$. 
\end{theorem}
\begin{proof}
As recalled above, since $\Shell_{{\boldsymbol{\mathit{Op}}}} (\rho)$
is complete for $\boldsymbol{\mathit{Op}}$, we have that
$\Shell_{{\boldsymbol{\mathit{Op}}}} (\rho)$ is complete for 
$\mathfrak{L}_{\mathit{Op}}$. Moreover, 
$\Shell_{{\boldsymbol{\mathit{Op}}}} (\rho)\sqsubseteq \rho$.
Let us consider any $\eta \in \uco(\wp(\mathbb{T})_\supseteq)$ such that 
$\eta\sqsubseteq \rho$ and $\eta$ is complete for 
$\mathfrak{L}_{\mathit{Op}}$. Let us check that $\eta$ is complete for 
$\boldsymbol{\mathit{Op}}$. Consider $\boldsymbol{\mathit{op}} \in 
\boldsymbol{\mathit{Op}}$ and, for simplicity, assume that
$\boldsymbol{\mathit{op}}$ is unary. 
Given $T \in \wp(\mathbb{T})$,  we consider an environment $\xi \in
\mathbb{E}$ such that $\xi(X)=T$.  Hence, by completeness of $\eta$
for $\mathfrak{L}_{\mathit{Op}}$, 
we have that $\eta
(\boldsymbol{\mathit{op}} (T)) = \eta
(\boldsymbol{\mathit{op}} (\xi(X))) =
\ok{\eta (\grass{\mathit{op}(X)}\xi)} =
\ok{\grass{\mathit{op}(X)}^\eta \dot{\eta} (\xi)} = \eta 
(\boldsymbol{\mathit{op}} ( \eta (\xi (X)))) = 
\eta 
(\boldsymbol{\mathit{op}} ( \eta (T)))$.  
Therefore, $\eta \sqsubseteq \Shell_{{\boldsymbol{\mathit{Op}}}}
(\rho)$. This implies that $\Shell_{\mathfrak{L}_{\mathit{Op}}}
(\rho)$ exists and  $\Shell_{\mathfrak{L}_{\mathit{Op}}} (\rho) =
\Shell_{{\boldsymbol{\mathit{Op}}}} (\rho)$. 
\end{proof}

Obviously, an analogous result  holds for complete cores as well. 
This general result can be applied to the $\mus$-calculus. 
Recall that $\mathrm{TT}$ denotes the set  of trace
transformers of the $\mus$-calculus, where the universal quantifier is
restricted to a unary operator. Let us denote by 
$\mathit{TT}$ the corresponding set of temporal
connectives of the  $\mus$-calculus so that 
$\mathfrak{L}_{\mathit{TT}} \subseteq \clm$ is a slight restriction of
the $\mus$-calculus 
where universal quantifications are unary. 
Consider any set  $\mathit{Op} \subseteq \mathit{TT}$ of temporal
connectives, that gives rise to the language 
$\mathfrak{L}_{\mathit{Op}} \subseteq \mathfrak{L}_{\mathit{TT}}$,  and assume that the
complete shell 
$\Shell_{{\boldsymbol{\mathit{Op}}}} (\rfm)$ of the universal closure $\rfm$
for the trace transformers in $\boldsymbol{\mathit{Op}}$ exists. Then,
by Theorem~\ref{teofix}, it turns out that $\Shell_{\mathfrak{L}_{\mathit{Op}}} (\rfm) =
\Shell_{{\boldsymbol{\mathit{Op}}}} (\rfm)$. Analogously, this also
holds for complete cores.  
Consequently, as far as the core is concerned, we have that
$$\ok{\core_{\mathfrak{L}_{\mathit{TT}}} (\rho^\forall_{M})} = 
\lambda X. \varnothing.$$
On the other hand, by Theorem~\ref{lconcbis}, it turns out that
$$\ok{\shell_{\mathfrak{L}_{\mathit{TT}}} (\rho^\forall_{M})} = 
\lambda X. X\cap M^*.$$

Thus, in general, in order to obtain the complete shell/core of the
universal closure for some fragment $\mathfrak{L}_{\mathit{Op}}$ of
the $\mus$-calculus it is enough to characterize the complete
shell/core for the corresponding set ${\boldsymbol{\mathit{Op}}}$ of
trace transformers. For example, if $\mathit{Op}$ includes arbitrary
disjunction but does not include time reversal, so that
$\mathfrak{L}_{\mathit{Op}}$ is a future-time language, by the result
mentioned at the end of Section~\ref{atc}, we have that
$\ok{\shell_{\mathfrak{L}_{\mathit{Op}}} (\rho^\forall_{M})}=\lambda
X. X\cap M$.

\section{Conclusion}

This paper studied the completeness of state-based w.r.t.\ trace-based
model checking by using a body of techniques based on abstract
interpretation.  By using a slogan, this study showed that
``\textit{the state-based model checking is intrinsically incomplete
w.r.t.\ trace-based model checking}'', since no refinement or
abstraction of the standard state-based semantics for model checking
induced by the universal/existential abstraction of past- and
future-time specification languages can lead to a semantics whose
corresponding model checking is complete for the trace semantics of
the specification language.  

The results of this paper suggest some research directions.  An abstract
interpretation-based approach to model checking for modal Kripke
transition systems has been studied by Huth et al.~\cite{hjs01}. It is
then interesting to investigate whether the framework of modal
transition systems based on three-valued logics affects the
incompleteness of states w.r.t.\ traces.  In view of the
characterizations of transition systems provided by Theorem~\ref{inj}
and Corollary~\ref{corma}, it is also interesting to determine
fragments of $\mu$-calculi and classes of transition systems such that
the universal/existential abstraction results to be complete.
Finally, it is certainly interesting to investigate how completeness
of state-based abstractions interacts with the presence of spurious
counterexamples in abstract model checking.  The works by Clarke et
al.\ \cite{cgjlv00,cgjlv03,cjlv02} on spurious counterexamples
originated from the idea of systemically refining abstract models in
order to enhance their precision. A spurious counterexample is an
abstract trace which is an artificial counterexample generated by the
approximation of the abstract model checker, namely there exists a
concrete trace approximated by the spurious counterexample which is
not a real counterexample.  Clarke et al.\ devised a methodology for
refining an partition-based abstract model relatively to a given
temporal specification $\phi$ by using the spurious counterexamples
provided by the abstract model checker on $\phi$.  The relationship
between spurious counterexamples and the trace-semantics of temporal
calculi has not been investigated from an abstract
interpretation-based perspective and we believe that the results of
this paper might shed some light on these issues.

\paragraph*{{\it Acknowledgements.}}
We are grateful to the anonymous referees 
for their helpful comments.
This work is an extended and revised version of two conference papers
\cite{gr02,ran01} and
was partially supported by the FIRB Project RBAU018RCZ
``Abstract interpretation and model checking for the verification
of embedded systems''  and 
by the COFIN2004
Project ``AIDA: Abstract Interpretation Design and Applications''.

\end{document}